\documentclass[journal]{IEEEtran}

\usepackage{bm}
\usepackage[cmex10]{amsmath}
\usepackage{amssymb}
\usepackage[pdftex]{graphicx}
\usepackage{stfloats}
\usepackage{algorithm}
\usepackage{algorithmic}
\usepackage[numbers,sort&compress]{natbib}

\usepackage{color}
\ifCLASSINFOpdf

\else

\fi

\begin{document}

\title{Latency Minimization for Hybrid-Frequency UHD Upload in Double-IRS-Aided HSR Networks}

%
%
%
\author{Tianyou Li,
      Huawei Tong,
      Dapeng Li, ~\IEEEmembership{Member,~IEEE}
  
\thanks{This work was supported by grants from the National Natural Science Funds of China under Contract No. 61801240, 62371246 and 61701253. (Corresponding author: Dapeng Li email: dapengli@njupt.edu.cn) }
}

%
%

\markboth{Journal of \LaTeX}%
{Shell \MakeLowercase{\textit{et al.}}: Bare Demo of IEEEtran.cls for IEEE Journals}
%



\maketitle

\begin{abstract}
\small {Real-time mechanical fault diagnosis in high-speed railway (HSR) networks requires ultra-reliable and low-latency upload of ultra-high-definition (UHD) video streams. 
However, energy constraints of trackside cameras and severe transmission latency pose critical challenges. This paper proposes a novel 6G infrastructure-to-vehicle (I2V) architecture employing double intelligent reflecting surfaces (IRSs) to enhance wireless powered communication network (WPCN) and hybrid-frequency data transmission. 
Crucially, to guarantee the quality of experience (QoE) for in-cabin passengers using Mobile Multimedia Broadcasting Services (MBMS), a strict zero-forcing spatial interference isolation constraint is imposed via the window-mounted IRS. 
We formulate a weighted latency minimization problem and develop a block coordinate descent (BCD) algorithm. 
Downlink energy beamforming and uplink information transmission are alternately optimized utilizing difference of convex (DCA) and semi-definite relaxation (SDR) techniques. 
Additionally, a low-complexity heuristic algorithm is proposed to mitigate the severe Doppler spread induced by train mobility. 
Simulation results demonstrate that the proposed scheme significantly reduces upload latency to meet stringent URLLC thresholds while ensuring interference isolation within the carriage.}
\end{abstract}


%
\IEEEpeerreviewmaketitle

\section{Introduction}

\IEEEPARstart{R}{ECENTLY}, the investigation of high-speed railway (HSR) has attracted a lot of interest since an increasing number of people choose the convenient and economical way to travel to another place.
With the rapid development of HSR, railway safety has attracted a great deal of attention. 
In current HSR operations, trackside fault detection systems (TFDS) play a vital role in capturing millimeter-level mechanical defects of passing trains \cite{11345161}.
To enable real-time monitoring of the train condition, a large number of Ultra-high-definition (UHD) multimedia sensors (such as UHD cameras and collaborative sensing nodes) have been deployed along the track.
However, traditional architectures route these massive UHD diagnostic images to remote ground servers, incurring catastrophic minute-level latency.
To prevent imminent derailments, it is imperative to establish a 6G Infrastructure-to-Vehicle (I2V) direct communication architecture.
These edge nodes are required to upload captured UHD surveillance video streams directly to the passing train in real time, supporting immediate decision-making by the Automatic Train Control (ATC) system and ensure operational safety \cite{10962291}.

To meet the increasing capacity demands of railway UHD surveillance video streams, the spectrum extension of 5G communications is envisaged at higher frequency bands with broader available spectrum, including frequency bands higher than 6 GHz, up to 300 GHz, since most of the spectrum sub-6 GHz is already utilized currently. 
Some researchers proposed the heterogeneous network to solve the issue. Lower frequency bands, such as the existing cellular band of macro cells, are used to provide basic coverage while higher frequency bands in small cells are used to provide high-speed data transmission \cite{11408918, 10198464}. 
Consequently, it is widely recognized that high-frequency and low-frequency systems will coexist within railway communication architectures for the foreseeable future.

However, the HSR multimedia sensing network still has many other issues that need to be overcome.
With the development of railroad cameras, a large number of UHD cameras deployed along the trackside make the manual management of the battery very costly \cite{10938652}.
Furthermore, deploying wired connections for these trackside cameras is cost-prohibitive and impractical.
Hence, wireless power technology such as radio frequency energy harvesting has emerged as a highly promising solution to power the cameras \cite{11079660}.
While WPCN effectively mitigates the energy bottleneck for these edge sensing nodes, the massive data volume of UHD video transmission imposes another formidable challenge: extreme communication latency.
As intelligent railway systems continue to evolve, existing low-rate communication systems can no longer meet the transmission requirements for UHD video streams. 
During the extremely short, proximity-based communication window when the train passes the trackside cameras at high speed, these cooperative sensing nodes are triggered to execute an opportunistic burst transmission of UHD diagnostic streams to the train.
Due to the limited computational resources at these edge nodes, complex video compression algorithms are often infeasible \cite{10496847}.
According to prevailing UHD multimedia transmission specifications, the end-to-end latency budget for mission-critical interactive video streams is subject to stringent millisecond-level thresholds. 
Exceeding these thresholds not only leads to media playback stalling, but also severely compromises the safety of the ATC system.
\begin{figure}{}
\centering
  \includegraphics[width=3.1in,height=2in]{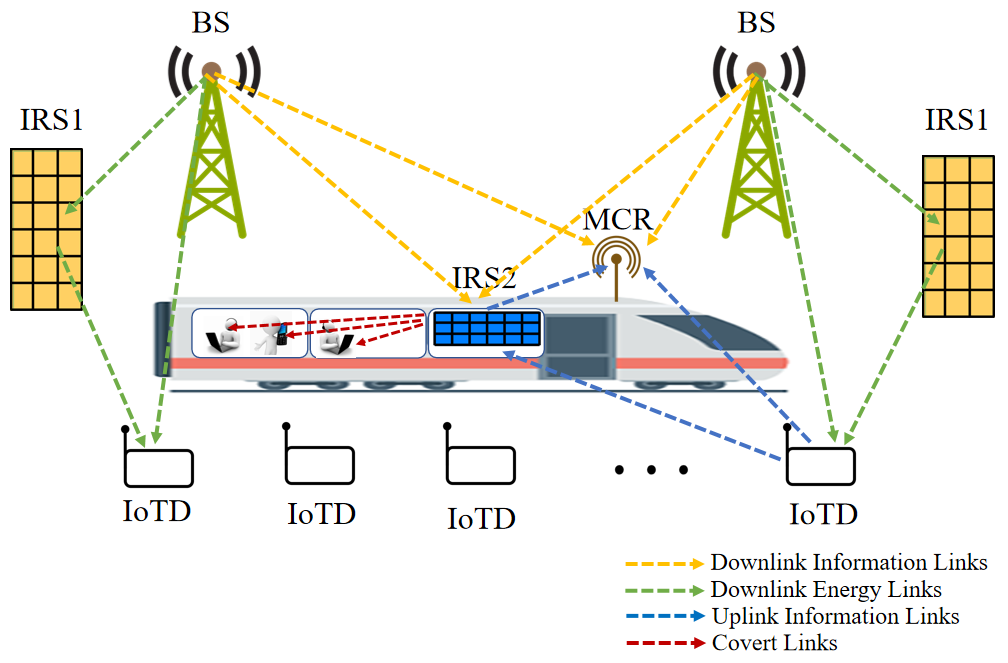}
  \caption{Double-IRSs aided massive MIMO communication system in complex high-speed railway traveling environment. In each communication module, there are two IRSs and the base station with the IRS controller via the wireless backhaul link.}\label{system}
\end{figure}

To address these challenges, Intelligent Reflecting Surface (IRS) technology stands out for its ability to reconfigure the wireless propagation environment \cite{10645688}. 
Many researchers have investigated the fundamental properties of IRS and its applications in various systems. 
Most recently, some initial efforts have been devoted to the high-speed scenarios aided by IRS.
The applications of IRS in HSR communication system were investigated in \cite{10287590}.
\cite{HSRIRS} proposed a deep reinforcement learning method, which makes real-time decision-making of beamforming truly viable in the dynamic HSR network.
Moreover, the authors in \cite{HSRIRS5} solved the sum rate maximization problem of multicell MIMO communication systems aided by IRS, which also applies to mobile scenarios.
However, different from our work, it should be emphasized that all works above have been dedicated to maximizing the spectral efficiency of communication system.
They all considered using IRS to optimize the communications from the BS to all users.
\cite{CHET1,10816075} considered channel estimation problem of IRS aided HSR system.
They proposed low overhead channel estimation algorithms for high-speed scenarios.
However, they didn't consider using IRS to improve performance of HSR system.
\cite{11175020} considered use IRS to suppress interference.
They optimize the phase shifts to enhance the HSR network capacity against the interference.
But they didn't consider the impact of IRS on information latency, nor did they consider collaboration between multiple frequency bands.
While aforementioned research has explored the application of IRS in HSR scenarios, most studies have focused on improving spectral efficiency, neglecting the stringent latency constraints of multimedia transmission and the potential for collaboration between heterogeneous frequency bands.

In this paper, we examine a WPCN to transfer power to the HSR multimedia sensing network aided by double IRS.
We consider the cooperation in different frequency bands for performance improvement.
Given that passengers in the train are enjoying Mobile Multimedia Broadcasting Services (MBMS), we have introduced a zero-forcing interference isolation constraint.
By precisely designing the phase array of the IRS2, we ensure that the uplink monitoring signal is physically isolated from the in-train entertainment broadcast channels in space.
Our main contributions are summarized in the following:
\begin{itemize}
\item We propose a novel double IRS-aided architecture to enhance the WPCN in HSR multimedia sensing networks.
We deploy double IRSs, with one deployed along the track and the other one is deployed as an electromagnetically transmissive surface on the train windows.
The trackside surfaces are used to assist in energy transfer, while the window surfaces assist in the high-rate upload of multimedia video streams.
This system paradigm facilitates the flexible deployment of trackside cameras and also enhances the battery life of cameras.
\item Moreover, we investigate and formulate the weighted latency minimization problem of video stream uplink transmission as the multimedia quality of service (QoS) optimization model. 
By introducing slack variables, we decouple the latency minimization problem into two sub-problems corresponding to the downlink energy transfer and uplink video streams upload phases.
We reformulate the original problem into a more tractable form and propose to use the difference of convex algorithm to solve the energy beamforming matrix optimization in the downlink phase.
\item Next, in the uplink phase, we consider the hybrid high and low frequency bands communication system, which can improve the reliability of the system.
IRS2 reflects the incident signals of both high and low frequency bands, and achieves the local optimality in both frequency bands simultaneously.
Furthermore, by precisely optimizing the phase shifts of IRS2, we formulate a zero-forcing spatial isolation constraint. 
This effectively prevents the uplink surveillance signals from causing co-channel interference to the downlink MBMS enjoyed by passengers inside the carriage.
We reformulate the original problem into the Semi-definite Relaxation (SDR) problem and relax the rank one constraint to solve it.
\item Based on the optimized phase shifts by the proposed algorithm, they can be further adjusted by a low complexity heuristic algorithm to mitigate the Doppler spread.
This critical operation artificially extends the channel coherence time, thereby preserving the effective system bandwidth required for continuous UHD multimedia delivery and strictly preventing latency explosions.
\item Finally, we provide a large number of numerical results to verify the effectiveness of our proposed hybrid transmission scheme and Doppler mitigation algorithm.
\end{itemize}

\textcolor{black}{\emph{Organization}: The rest of the paper is organized as follows.
The system model and problem formulation are discussed in Section II. Our proposed algorithm is presented in Section III, IV and V. 
Results are described in Section VI and the conclusions are drawn in Section VII.}

\emph{Notations}: Bold lower and upper case letters denote vectors and matrices respectively.
$\mathrm{Tr}(\cdot)$ denotes the trace operator. $\mathbb{C}^{m\times n}$ denotes the space of complex matrices of dimensions given as in the superscripts $m$ and $n$.  
For a general matrix $\textbf{A}$, 
$\textbf{A}_{(i,j)}$ denotes the $(i, j)$-th element.
The superscript $(\cdot)^{T}, (\cdot)^\mathrm{*}, (\cdot)^{H}$ and $(\cdot)^{-1}$ stand for transpose, complex conjugate, Hermitian transpose and matrix inverse, respectively. 
$\odot$ denotes the dot product. 
$\mathbb{E}\{\cdot\}$ stands for the statistical expectation. 
 $x\sim{U}(a,b)$ represents a random variable following the uniform distribution between $a$ and $b$. $x\sim\mathcal{CN}(0,\sigma^2)$ represents a random variable following the distribution of circularly symmetric complex Gaussian with zero mean and variance $\sigma^2$. 
\textcolor{black}{$\mathcal{O}(\cdot)$ denotes the standard big-O notation.}

\section{System Model and Problem Formulation}
In this section, we elaborate on the system model from both downlink energy transmission phase and uplink information transmission phase.
As shown in Fig. \ref{system}, a wireless power HSR multimedia sensing network aided by double IRSs is considered in our work.
\textcolor{black}{For the sake of simplicity of problem formulation, we don’t consider the BS/device handover problem when train travels across cells.}
Subsequently, a latency-minimization problem is formulated for our IRS-aided HSR multimedia sensing network.

\begin{figure}
  \begin{center}
  \includegraphics[width=3.5in,height=2in]{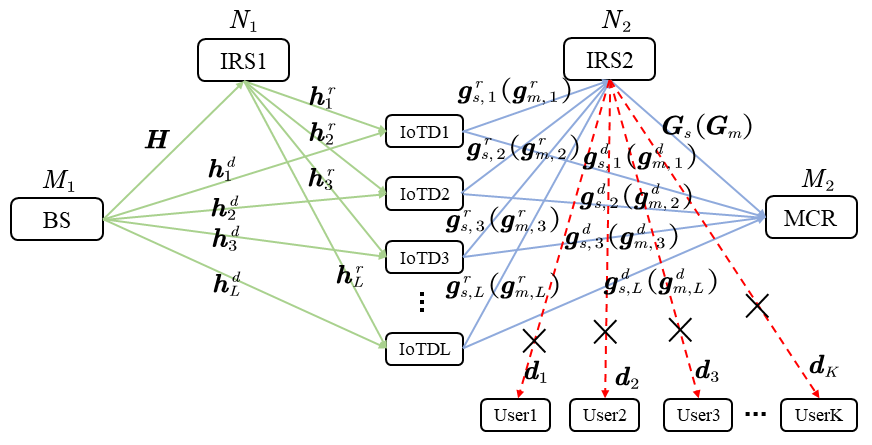}
  \caption{The channel model shown in the figure is considered. In energy transmission phase, no matter what the frequency band of the signal can transmit energy, while in information transmission phase, the signal in the same frequency band is treated as noise. }\label{systemparadiam}
  \end{center}
\end{figure}
\subsection{Energy Transmission Model}
Considering a practical downlink energy transmission scenario consisting of one energy transmission BS, one IRS1, $L$ single antenna cameras that are deployed along the trackside to capture diagnostic video streams. 
The set of cameras is defined as $\mathcal{L}=\left\{1,2,\cdots,L\right\}$.
The number of antennas at the BS and passive reflecting elements at the IRS1 are denoted by $M_1$ and $N_1$, respectively. 
The set of passive elements in IRS1 is defined as $\mathcal{N}_1=\left\{1,2,\cdots,N_1\right\}$.
For flexibility and scalability of deployment, in this phase, the BS needs to exploit wireless power transfer to $L$ cameras aided by IRS1.
The downlink baseband channels from the BS to the IRS1, the IRS1 to the $l$-th camera, the BS to the $l$-th camera are denoted by $\bm{H}\in\mathbb{C}^{N_1\times M_1}$, $\bm{h}_l^r\in\mathbb{C}^{1\times N_1}$ and $\bm{h}_l^d\in\mathbb{C}^{1\times M_1}$, respectively.
By beneficially introducing the different phase shifts to the incident signals, the IRS1 can make them added constructively at all cameras to enhance the energy transfer efficiency.
Let $\theta_{1,n_1}\in[0,2\pi]$ be the phase shift introduced to the incident signals passing the $n_1$-th passive element by the IRS1.  
As for the amplitude reflection coefficient of the IRS1,
we simply set it to 1 for all reflection elements, which has been proved to be optimal in \cite{11005674}.
In the following, a synchronous time-splitting protocol is adopted in our work.
Specifically, in the first $\beta T\ (0<\beta<1)$ amount of time during each unit time, the BS transmits energy to all cameras.
In the second $(1-\beta)T$ amount of time, all cameras upload the information to the MCR for decision-making by ATC.
The time block $T$ is normalized to unit.

\textcolor{black}{Generating multiple beams can help improve the fairness of energy harvesting when there are multiple cameras \cite{mbc1}.}
Therefore, we assume that the fully-connected digital beamforming structure is deployed at the BS, so it can broadcast energy to all cameras through $M_1$ energy beams.
The beam signal carrying energy can be presented as $\bm{x}_d=\sum^{M_1}_{i=1}\bm{w}_is_i$, 
where $\bm{w}_i\in\mathbb{C}^{M_1\times1}$ denotes the $i$-th downlink energy beam of $\bm{W}$ and $s_i$ denotes the energy carrying signal, which is assumed to be an independent identically distributed random variable with zero mean and unit variance.
Therefore, the transmit power of the BS can be represented as $p_t=\mathbb{E}[\|\bm{x}_d\|^2]=\sum^{M_1}_{i=1}\|w_i\|^2$.
Then, all cameras can receive the direct energy from the BS and the reflecting energy from the IRS1.
Upon denoting the additive white Gaussian noise by $n\sim\mathcal{CN}(0,\sigma^2)$, the signal received at the $l$-th camera is readily formulated as
\begin{align}
y_{DL,l}=(\bm{h}^r_l\bm{\Phi}_1\bm{H}+\bm{h}_l^d)\bm{x}_d+n,
\end{align}
where $\bm{\Phi}_1=\text{diag}\{e^{j\theta_{1,1}},\dots,e^{j\theta_{1,n_1}},\dots,e^{j\theta_{1,N_1}}\}$ denotes the diagonal reflection-coefficients matrix.
By defining the equivalent channel as $\bar{\bm{h}}_l\triangleq\bm{h}^r_l\bm{\Phi}_1\bm{H}+\bm{h}_l^d$, the energy received by the $l$-th camera in the energy transmission phase can be expressed as
\begin{align}
E_l=\xi\beta\mathbb{E}\left[|y_{DL,l}|^2\right]=\xi\beta\sum^{M_1}_{i=1}\bm{\bar{h}}_l\bm{w}_i\bm{w}_i^H\bm{\bar{h}}_l^H,
\end{align}
where $0<\xi\leq1$ is the energy transmission efficiency from the BS to cameras.
By normalizing the uplink information transmission time, the transmitted power of the $l$-th camera can be presented as
\begin{align}
p_l=\frac{E_l}{1-\beta}=\frac{\xi\beta\sum^{M_1}_{i=1}\bm{\bar{h}}_l\bm{w}_i\bm{w}_i^H\bm{\bar{h}}_l^H}{1-\beta}.\label{p_l}
\end{align}
\subsection{Information Transmission Model}
Next, in the uplink information transmission phase, each camera needs to report information to the MCR, which is equipped with $M_2$ antennas.
Then, the MCR transmits the information to the driver's compartment via wired transmission.
Since the performance bottleneck of the second uplink phase lies in the wireless transmission part, we only focus on the hop from the cameras to the MCR.
The IRS2 deployed on the train windows, which is equipped with $N_2$ passive elements, can be used to enhance the reflected signals from the cameras to the MCR.
The set of passive elements in IRS2 is defined as $\mathcal{N}_2=\left\{1,2,\cdots,N_2\right\}$.
We denote the phase shifts of the $n_2$-th element of IRS2 as $\theta_{2,n_2}\in[0,2\pi]$.
Our proposed HSR multimedia sensing network can work in hybrid frequency band. 
Considering carrier aggregation technology, all cameras can choose to operate in both the sub-6 GHz with carrier frequency $f_s$ and mmWave frequency band $f_m$ to upload the information.
Specifically, the report information can be represented as
\begin{align}
x_{\mathcal{F},l}=\sqrt{p_{\mathcal{F},l}}s_{\mathcal{F},l}, \ \mathcal{F}\in\{s,m\},    
\end{align}
where $s_{\mathcal{F},l}$ denotes the report information transmitted by the
$l$-th camera to the MCR, which satisfies $\mathbb{E}[|s_{\mathcal{F},l}|^2] = 1$ and $\mathbb{E}[s_{\mathcal{F},k}s^*_{\mathcal{F},l}] = 0$, for $l\neq k$.
And by defining the transmit power allocation factor $\bm{\gamma}=[\gamma_1, \gamma_2,\dots,\gamma_L]$, we have $p_{s,l}=\gamma_lp_{l}$, $p_{m,l}=(1-\gamma_l)p_{l}$.
The IRS2 deployed on the window is the transparent IRS whose copper backplane is removed \cite{11151987}.
It can work in transmissive type.

For sub-6 or mmWave band, the received signal at the MCR in the UL phase can be expressed as
\begin{align}
\bm{y}_{\mathcal{F},UL}=\sum_{l=1}^L\left(\sqrt{p_{\mathcal{F},l}}\left(\bm{G}_\mathcal{F}\bm{\Phi}_2\bm{g}_{\mathcal{F},l}^r+\bm{g}^d_{\mathcal{F},l}\right)s_{\mathcal{F},l}\right)+\bm{n}_\mathcal{F},
\end{align}
where $\bm{n}_\mathcal{F}=[n_{\mathcal{F},1},n_{\mathcal{F},2},\dots,n_{\mathcal{F},M_2}]$ for $n_{\mathcal{F},m_2}\sim\mathcal{CN}(0,\sigma_\mathcal{F}^2),\ m_2=1,2,\dots,M_2$ and  $\bm{\Phi}_2=\text{diag}\{e^{j\theta_{2,1}},\dots,e^{j\theta_{2,n_2}},\dots,e^{j\theta_{2,N_2}}\}$.
The uplink baseband channels from the IRS2 to the MCR, the $l$-th camera to the IRS2, the $l$-th camera to the MCR are denoted by $\bm{G}_{\mathcal{F}}\in\mathbb{C}^{M_2\times N_2}$, $\bm{g}_{\mathcal{F},l}^r\in\mathbb{C}^{N_2\times 1}$ and $\bm{g}_{\mathcal{F},l}^d\in\mathbb{C}^{M_2\times 1}$, respectively.
Then, at the receiver, the \textcolor{black}{linear} multi-user decoder technique is invoked to reduce the complexity.
\textcolor{black}{$\bm{F}_{\mathcal{F}}=[\bm{f}_{\mathcal{F},1},\bm{f}_{\mathcal{F},2},\dots,\bm{f}_{\mathcal{F},L}]$ denotes the multi-user decoder matrix.}
As for the $l$-th camera, the recovered signal is presented as
\begin{align}
\hat{s}_{\mathcal{F},l}=\bm{f}_{\mathcal{F},l}^H\left[\sum_{i=1}^L\left(\sqrt{p_{\mathcal{F},i}}\left(\bm{G}_\mathcal{F}\bm{\Phi}_2\bm{g}_{\mathcal{F},i}^r+\bm{g}^d_{\mathcal{F},i}\right)s_{\mathcal{F},i}\right)+\bm{n}_{\mathcal{F}}\right].
\end{align}
Then, the SINR of the $l$-th camera is presented as 
\begin{equation}
    \text{SINR}_{\mathcal{F},l}=\frac{p_{\mathcal{F},l}\bm{f}_{\mathcal{F},l}^H\bar{\bm{g}}_{\mathcal{F},l}\bar{\bm{g}}_{\mathcal{F},l}^H\bm{f}_{\mathcal{F},l}}
    {\bm{f}_{\mathcal{F},l}^H\left(\sum\limits_{i=1,i\neq l}^L p_{\mathcal{F},i}\bar{\bm{g}}_{\mathcal{F},i}\bar{\bm{g}}_{\mathcal{F},i}^H+\sigma_\mathcal{F}^2\bm{I}\right)\bm{f}_{\mathcal{F},l}}\label{SINR_F},
\end{equation}
so the capacity of the channel for $l$-th camera can be expressed as $C_{\mathcal{F},l}=(1-\beta)B_\mathcal{F}\log_2\left[1+\text{SINR}_{\mathcal{F},l}\right]$,
where $p_{s,l}+p_{m,l}=p_{l}$ and $\bar{\bm{g}}_{\mathcal{F},l}\triangleq\left(\bm{G}_\mathcal{F}\bm{\Phi}_2\bm{g}_{\mathcal{F},l}^r+\bm{g}^d_{\mathcal{F},l}\right)$ for $\mathcal{F}\in\left\{s,m\right\}$.  
It is worth noting that passengers inside the train carriages are typically served by the downlink MBMS via the onboard access points. 
Since the I2V uplink diagnostic video transmission shares the same frequency band with the in-cabin MBMS to improve spectral efficiency, the high-rate uplink signals penetrating through the window IRS2 will cause severe co-channel interference to the passengers' transmission.
To guarantee the Quality of Experience (QoE) of the MBMS, a stringent spatial interference isolation constraint must be imposed. 
Let $\mathcal{K}=\left\{1,2,\cdots,K\right\}$ denote the set of passengers.
Assuming that the metallic body of the HSR carriage can severely attenuate direct external radio waves, the cascaded channel penetrating through the window-mounted IRS dominates the interference.
Therefore, the interference isolation constraint can be formulated as:
\begin{equation}
\sum_{k=1}^{K}\sum_{l=1}^{L}\left|\bm{d}_{\mathcal{F},k}\bm{\Phi}_2\bm{g}_{\mathcal{F},l}^r\right|^2=0, \label{zeroforce}
\end{equation}
where $\bm{d}_{\mathcal{F},k}\in\mathbb{C}^{1\times N_2}$ denotes the baseband channel from the IRS2 to the $k$-th passenger.
In our work, we assume that all CSI are perfectly known.
Let $v_l$ denote the data volume of the UHD video chunk generated by the $l$-th camera. 
To smooth the extreme traffic burstiness of H.265/HEVC encoding while avoiding queueing backlog, we adopt a time-based chunking mechanism with a strict buffering window $T_{chunk}$. 
Consequently, $v_l$ represents the accumulated 4K UHD payload within $T_{chunk}$.
To prevent buffer overflow and guarantee the continuous streaming of the mission-critical diagnostic video, we need to minimize the total upload latency for all cameras to satisfy the queuing stability conditions.
Upon defining the volume allocation factor $\bm{\alpha}=[\alpha_1,\alpha_2,\dots,\alpha_L]^T$, from presented above, the information upload latency of the $l$-th camera can be readily calculated by selecting the maximum value between those transmitted in the sub-6 GHz or the mmWave band, formulated as
\begin{equation}
D_l(\bm{\alpha},\beta,\bm{W},\bm{\theta}_1,\bm{\gamma},\bm{F},\bm{\theta}_2)=\max\left\{\frac{\alpha_l v_l}{C_{s,l}},\frac{(1-\alpha_l)v_l}{C_{m,l}}\right\}\label{D_l}.
\end{equation}
\subsection{Problem Formulation}
In this work, we aim for minimizing sum weighted information upload latency of all cameras, by jointly optimizing variables $\beta$, $\bm{\alpha}$, $\bm{W}$, $\bm{\theta}_1$, $\bm{p}_s$, $\bm{F}_{\mathcal{F}}$ and $\bm{\theta}_2$. 
Specifically, upon defining the weight of the $l$-th camera as $\varpi_l$, the weighted latency minimization problem is
formulated as
\begin{align}
\mathcal{P}0:&\min_{\bm{\alpha},\beta,\bm{W},\bm{\theta}_1,\bm{\gamma},\bm{F},\bm{\theta}_2}\sum_{l=1}^{L}\varpi_lD_l\label{P1}\\
\text{s.t.}\ &0<\beta<1,\tag{\ref{P1}{a}}\label{P1a}\\
&0\leq\alpha_l,\gamma_l\leq1, \quad \forall{l}\in\mathcal{L},\tag{\ref{P1}{b}}\label{P1b}\\
&0\leq\theta_{1,n}<2\pi,  \quad \forall{n}\in\mathcal{N}_1\tag{\ref{P1}{c}}\label{P1c},\\
&0\leq\theta_{2,n}<2\pi,  \quad \forall{n}\in\mathcal{N}_2\tag{\ref{P1}{d}}\label{P1d},\\
&\lVert\bm{W}\rVert_F^2\leq P_{\text{max}}\tag{\ref{P1}{e}}\label{P1e},\\
&\sum_{k=1}^{K}\sum_{l=1}^{L}\left|\bm{d}_{\mathcal{F},k}\bm{\Phi}_2\bm{g}_{\mathcal{F},l}^r\right|^2=0\tag{\ref{P1}{f}}\label{P1f}.
\end{align}
(\ref{P1a}) and (\ref{P1b}) are allocation factor constraints.
When $\alpha_l=1$, it means that the channel conditions in the mmWave band are extremely poor and even blocked for the $l$-th camera.
At this time, the information should be all transmitted through the sub-6 band.
When $\alpha_l=0$ is the opposite; 
(\ref{P1c}) and (\ref{P1d}) specifies the range of the phase shifts of the passive elements deployed on the double IRSs. 
(\ref{P1e}) restrict the total transmit power at the BS side.
Finally, (\ref{P1f}) represents the zero-forcing spatial interference isolation constraint to protect the $K$ MBMS users in the carriage from uplink UHD video transmissions.

The existing schemes are not applicable to solving the proposed problem due to the complex coupling effect of so many variables.
It is a challenging problem to obtain a globally optimal solution directly. 
Consequently, we proposed a low \textcolor{black}{complexity} locally suboptimal solution to handle the proposed optimization problem, which will be described in detail in the next section.
\section{Optimization for Downlink Energy Transmission}
As mentioned above, owing to the coupling effect of many variables, the information upload weighted latency minimization problem cannot be solved directly.
By reformulating the original problem into the more tractable form, we can decouple the problem into two subproblems, downlink energy optimization and uplink information optimization. 
Then, the formulated problem can be solved by alternately optimizing these subproblems.
In this section, we firstly focus our attention on the optimization of downlink energy transmission with uplink setting fixed.
Specifically, we firstly optimize the transmission volume allocation factor $\bm{\alpha}$.

For convenient analysis, we define $v_{s,l}\triangleq \alpha_l v_l$.
Given $\beta,\bm{W},\bm{\theta}_1,\bm{\gamma},\bm{F},\bm{\theta}_2$, from (\ref{D_l}), the information upload latency of the $l$-th camera can be reformulated as
\begin{equation}
    D_l(\bm{\alpha})=\left\{
\begin{aligned}
&  \frac{v_l-v_{s,l}}{C_{m,l}}, \ 0\leq v_{s,l}\leq\frac{v_lC_{s,l}}{C_{s,l}+C_{m,l}},\\
&  \frac{ v_{s,l}}{C_{s,l}},\quad\quad \frac{v_lC_{s,l}}{C_{s,l}+C_{m,l}}<v_{s,l}\leq v_l.
\end{aligned}
\right.
\end{equation}
It's obviously that $D_l(\bm{\alpha})$ decreases upon increasing $v_{s,l}$ in the range of $v_{s,l}\in\left[0,\frac{v_lC_{s,l}}{C_{s,l}+C_{m,l}}\right]$.
And $D_l(\bm{\alpha})$ increases upon increasing $v_{s,l}$ in the range of $v_{s,l}\in\left(\frac{v_lC_{s,l}}{C_{s,l}+C_{m,l}},v_l\right]$.
Consequently, $D_l(\bm{\alpha})$ can achieve the minimum value when 
\begin{equation}
    v^{\star}_{s,l}=\frac{v_lC_{s,l}}{C_{s,l}+C_{m,l}}.\label{v_sl}
\end{equation}
Substituting (\ref{v_sl}) into (\ref{P1}), we can reformulate the objective function $\mathcal{P}0$ as 
\begin{align}
\mathcal{P}\text{1}:&\min_{\beta,\bm{W},\bm{\theta}_1,\bm{\gamma},\bm{F},\bm{\theta}_2}\sum_{l=1}^{L}\frac{\varpi_lv_{l}}{C_{s,l}+C_{m,l}}\label{P1_E}\\
&\text{s.t.}\quad (\ref{P1a}),(\ref{P1c}),(\ref{P1e})\tag{\ref{P1_E}{a}}\label{p1_Ea}.
\end{align}
Without loss of optimality, problem $\mathcal{P}1$ can be equivalently reformulated as the following optimization problem:
\begin{align}
\mathcal{P}\text{1-}E1:
&\min_{\beta,\bm{W},\bm{\theta}_1,\bm{\gamma},\bm{F},\bm{\theta}_2,\bm{\eta}}\sum_{l=1}^{L}\eta_l\label{P1_E1}\\
\text{s.t.}\quad &\frac{\varpi_lv_{l}}{C_{s,l}+C_{m,l}}\leq\eta_l,\quad \forall{l}\in\mathcal{L}\tag{\ref{P1_E1}{a}}\label{P1_E1a},\\
&(\ref{P1a}),(\ref{P1c}),(\ref{P1e})\tag{\ref{P1_E1}{b}}\label{p1_E1b}.
\end{align}
To facilitate the derivation of the optimal conditions, the partial Lagrangian function associated with problem $\mathcal{P}\text{1-}E1$ is formulated as
\begin{align}
\mathcal{L}(\bm{C}_{\mathcal{F}},\bm{\lambda},\bm{\eta})=\sum_{l=1}^L\eta_l+\sum_{l=1}^L\lambda_l\left[\varpi_lv_{l}-\eta_l\left(C_{s,l}+C_{m,l}\right)\right].
\end{align}
If $(\bm{C}^*_{\mathcal{F}},\bm{\eta}^*)$ is the solution of the problem $\mathcal{P}\text{1-}E1$, there must exist $\bm{\lambda}^*$ such that the following Karush-Kuhn-Tucker (KKT) conditions are satisfied
\begin{align}
&-\lambda_l^{\star}\eta_l^{\star}(\nabla C^{\star}_{s,l}+\nabla C^{\star}_{m,l})=0\label{KKT1},\quad {l}\in\mathcal{L},\\
&\frac{\partial\mathcal{L}}{\partial\eta_l}=1-\lambda_l^{\star}(C^{\star}_{s,l}+C^{\star}_{m,l})=0,\label{KKT2}\quad {l}\in\mathcal{L},\\
&\lambda_l^{\star}\left[\varpi_lv_{l}-\eta^{\star}_l\left(C^{\star}_{s,l}+C^{\star}_{m,l}\right)\right]=0,\label{KKT3}\quad {l}\in\mathcal{L},\\
&\lambda_l^{\star}\geq0,\label{KKT4}\quad {l}\in\mathcal{L},\\
&\varpi_lv_{l}-\eta^{\star}_l\left(C^{\star}_{s,l}+C^{\star}_{m,l}\right)\leq0\label{KKT5}\quad {l}\in\mathcal{L}.
\end{align}
From (\ref{KKT2}) and (\ref{KKT3}), we can easily obtain
\begin{align}
\lambda_l^{\star}=\frac{1}{C_{s,l}^{\star}+C_{m,l}^{\star}},\quad\eta_l^{\star}=\frac{\varpi_l v_l}{C_{s,l}^{\star}+C_{m,l}^{\star}},\label{inipara}
\end{align}
where $\bm{C}^{\star}_{\mathcal{F}}\triangleq\bm{C}_{\mathcal{F}}(\beta^{\star},\bm{W}^{\star},\bm{\theta}^{\star}_1,\bm{\gamma}^{\star},\bm{F}^{\star},\bm{\theta}_2^{\star})$ for $\mathcal{F}\in\{s,m\}$.
Then, with $\bm{\lambda}$ and $\bm{\eta}$ are initialized and fixed, we can reformulate the $\mathcal{P}\text{1-}E1$ as the following equivalent problem
\begin{align}
\mathcal{P}\text{1-}E2:
&\min_{\beta,\bm{W},\bm{\theta}_1,\bm{\gamma},\bm{F},\bm{\theta}_2}\sum_{l=1}^{L}\lambda_l\left[\varpi_lv_{l}-\eta_l\left(C_{s,l}+C_{m,l}\right)\right]\label{P1_E2}\\
\text{s.t.}&\quad(\ref{P1a}),(\ref{P1c}),(\ref{P1e})\tag{\ref{P1_E2}{a}}\label{p1_E2a}.
\end{align}
Through the aforementioned transformations, the original highly coupled sum-of-ratios problem is successfully converted into a mathematically tractable parametric form. 
In the subsequent subsections, we proceed to optimize the variables in the downlink energy transfer phase.
\subsection{Optimization of $\beta$}
Given $\bm{\alpha},\bm{W},\bm{\theta}_1,\bm{\gamma},\bm{F},\bm{\theta}_2$, problem $\mathcal{P}\text{1-}E2$ for optimizing time
allocation variable $\beta$ can be transformed as
\begin{align}
\mathcal{P}\text{2}:
&\max_{\beta}\sum_{l=1}^{L}\lambda_l\eta_l\left(C_{s,l}+C_{m,l}\right)\label{P2}\\
&\text{s.t.}\quad0<\beta<1\tag{\ref{P2}{a}}\label{p2a}.
\end{align}
Defining $a_{s,l}=\xi\sum_{i=1}^{M_1}\gamma_l\bar{\bm{h}}_l\bm{w}_i\bm{w}_i^H\bar{\bm{h}}_l^H\bm{f}_{s,l}^H\bar{\bm{g}}_{s,l}\bar{\bm{g}}_{s,l}^H\bm{f}_{s,l}$, $b_{s,l}=\\ \xi\sum_{j=1,j\neq l}^{L}\sum_{i=1}^{M_1}\gamma_j\bar{\bm{h}}_j\bm{w}_i\bm{w}_i^H\bar{\bm{h}}_j^H\bm{f}_{s,l}^H\bar{\bm{g}}_{s,j}\bar{\bm{g}}_{s,j}^H\bm{f}_{s,l}$, $c_{s,l}=\sigma_s^2\\\bm{f}_{s,l}^H\bm{f}_{s,l}$, $a_{m,l}=\xi\sum_{i=1}^{M_1}(1-\gamma_l)\bar{\bm{h}}_l\bm{w}_i\bm{w}_i^H\bar{\bm{h}}_l^H\bm{f}_{m,l}^H\bar{\bm{g}}_{m,l}\bar{\bm{g}}_{m,l}^H\bm{f}_{m,l}$, $b_{m,l}=\xi\sum_{j=1,j\neq l}^{L}
\sum_{i=1}^{M_1}(1-\gamma_j)\bar{\bm{h}}_j\bm{w}_i\bm{w}_i^H\bar{\bm{h}}_j^H\bm{f}_{m,l}^H\bar{\bm{g}}_{m,j}\bar{\bm{g}}_{m,j}^H\\\bm{f}_{m,l}$ and $c_{m,l}=\sigma_m^2\bm{f}_{m,l}^H\bm{f}_{m,l}$, respectively. 
The objective problem $\mathcal{P}\text{2}$ can be reformulated as
\begin{align}
\mathcal{P}\text{2-}E1:\max_{\beta}\sum_{l=1}^{L}\sum_{\mathcal{F}}^{\mathcal{F}\in\{s,m\}}f_{\mathcal{F},l}(\beta),
\end{align}
where $f_{\mathcal{F},l}(\beta)=\lambda_l\eta_lB_{\mathcal{F}}\left(1-\beta\right)\log_2\left(1+\frac{\beta a_{\mathcal{F},l}}{\left(b_{\mathcal{F},l}-c_{\mathcal{F},l}\right)\beta+c_{\mathcal{F},l}}\right)$ with 
$0<\beta<1$.
$\mathcal{P}\text{2-}E1$ is a convex optimization problem with respect to $\beta$.
The second order function of $\mathcal{P}\text{2-}E$ is strictly less than 0 in $0 < \beta < 1$.
Hence, a low complexity search method based on bisection is adopted to search the optimal $\beta^\star$ which satisfies $\frac{\partial(f_{\mathcal{F},l})}{\partial\beta}=0$.
\subsection{Joint Optimization of $\bm{W}$ and $\bm{\theta}_1$}
Given $\bm{\alpha},\beta,\bm{\gamma},\bm{F},\bm{\theta}_2$, problem $\mathcal{P}\text{1-}E2$ can be given as
\begin{align}
\mathcal{P}\text{3}:
&\max_{\bm{W},\bm{\theta}_1}\sum_{l=1}^{L}\sum_{\mathcal{F}}^{\mathcal{F}\in\{s,m\}}\lambda_l\eta_lC_{\mathcal{F},l}\label{P3}\\
&\text{s.t.}\ (\ref{P1c}), (\ref{P1e})\tag{\ref{P3}{a}}.
\end{align}
From \cite{MMSE}, the original problem can be reformulated as
\begin{align}
\mathcal{P}\text{3-}E1:
\min_{\bm{W},\bm{\theta}_1}\sum_{l=1}^{L}\sum_{\mathcal{F}}^{\mathcal{F}\in\{s,m\}}&\Big(\Gamma_{\mathcal{F},l}e_{\mathcal{F},l}(\bm{W},\bm{\theta}_1)-\lambda_l\eta_lB_{\mathcal{F}}\notag\\
-&\lambda_l\eta_lB_{\mathcal{F}}\log_2\left(\frac{\Gamma_{\mathcal{F},l}}{\lambda_l\eta_lB_{\mathcal{F}}}\right)\Big)\label{P3E}\\
\text{s.t.}\quad(\ref{P1c}), (\ref{P1e})\tag{\ref{P3E}{a}}
\end{align}
where ${\Gamma}_{\mathcal{F},l}$ is the newly introduced auxiliary variable and $e_{\mathcal{F},l}$ for $\mathcal{F}\in\{s,m\}$ represents the MSE of the $l$-th camera in sub-6 GHz or mmWave band, which is given by
\begin{align}
e_{\mathcal{F},l}\triangleq &\mathbb{E}\left[\left(\hat{s}_{\mathcal{F},l}-s_{\mathcal{F},l}\right) \left(\hat{s}_{\mathcal{F},l}-s_{\mathcal{F},l}\right)^H\right]\notag\\
=&\left(\sqrt{p_{\mathcal{F},l}}\bm{f}_{\mathcal{F},l}^H\bar{\bm{g}}_{\mathcal{F},l}-1\right)\left(\sqrt{p_{\mathcal{F},l}}\bm{f}_{\mathcal{F},l}^H\bar{\bm{g}}_{\mathcal{F},l}-1\right)^H\notag\\
&+\sum_{i\neq l}^Lp_{\mathcal{F},i}\bm{f}_{\mathcal{F},l}^H\bar{\bm{g}}_{\mathcal{F},i}\bar{\bm{g}}_{\mathcal{F},i}^H\bm{f}_{\mathcal{F},l}+\sigma_{\mathcal{F}}^2\bm{f}_{\mathcal{F},l}^H\bm{f}_{\mathcal{F},l}.\label{e_F}
\end{align}
\subsubsection{Optimization of $\Gamma_{\mathcal{F},l}$}
With given $\bm{W},\bm{\theta}_1$, we take the first order derivative of 
 $\mathcal{P}\text{3-}E1$ with respect to $\Gamma_{\mathcal{F},l}$ and set it equal to zero. 
The optimal $\Gamma_{\mathcal{F},l}^\star$ can be achieved by
\begin{align}
    \Gamma_{\mathcal{F},l}={\lambda_l\eta_l}B_{\mathcal{F}}e_{\mathcal{F},l}^{-1}.\label{opte}
\end{align}
\subsubsection{Optimization of $\bm{W}$}
With given $\Gamma_{\mathcal{F},l},\bm{\theta}_1$, $\mathcal{P}\text{3-}E$ can be rewritten as
\begin{align}
\mathcal{P}\text{3-}E2:
\min_{\bm{W}}\sum_{l=1}^{L}\sum_{\mathcal{F}}^{\mathcal{F}\in\{s,m\}}&\Gamma_{\mathcal{F},l}e_{\mathcal{F},l}(\bm{W})\label{P3E1}\\
\text{s.t.}\quad\lVert\bm{W}&\rVert_F^2\leq P_{\text{max}}.\tag{\ref{P3E1}{a}}\label{P3E1a}
\end{align}
The objective function $\mathcal{P}\text{3-}E2$ is nonconvex because of the nonconvex $e_{\mathcal{F},l}(\bm{W})$.
By defining $q_{\mathcal{F},l,i}\triangleq\bm{f}_{\mathcal{F},l}^H\bar{\bm{g}}_{\mathcal{F},i}$ and substituting (\ref{p_l}) into (\ref{e_F}), $e_{\mathcal{F},l}$ can be reformulated as
\begin{align}
e_{\mathcal{F},l}=&\sum_{i=1}^L\bar{\gamma}_{\mathcal{F},i}q_{\mathcal{F},l,i}q_{\mathcal{F},l,i}^H\|\bar{\bm{h}}_i\bm{W}\|^2+\sigma_{\mathcal{F}}^2\bm{f}_{\mathcal{F},l}^H\bm{f}_{\mathcal{F},l}\notag\\
&-\left(q_{\mathcal{F},l,l}+q_{\mathcal{F},l,l}^H\right)\sqrt{\bar{\gamma}_{\mathcal{F},l}}\|\bar{\bm{h}}_l\bm{W}\|+1,\label{e_F2}
\end{align}
where $\bar{\gamma}_{s,i}\triangleq\frac{\gamma_i\xi\beta}{1-\beta}$ and $\bar{\gamma}_{m,i}\triangleq\frac{(1-\gamma_i)\xi\beta}{1-\beta}$.
By substituting (\ref{e_F2}) into (\ref{P3E1}) and omitting the constant, we can reformulate the problem $\mathcal{P}\text{3-}E2$ as following:
\begin{align}
\mathcal{P}\text{3-}E3: \min_{\bm{W}}&\ f(\bm{W})=g(\bm{W})-h(\bm{W})\label{con_con}\\
\text{s.t.}&\quad\lVert\bm{W}\rVert_F^2\leq P_{\text{max}},\tag{\ref{con_con}{a}}\label{con_cona}
\end{align}
where
\begin{align}
g(\bm{W})&=\sum_{l=1}^{L}\sum_{\mathcal{F}}^{\mathcal{F}\in\{s,m\}}\sum_{i=1}^L\Gamma_{\mathcal{F},l}\bar{\gamma}_{\mathcal{F},i}q_{\mathcal{F},l,i}q_{\mathcal{F},l,i}^H\|\bar{\bm{h}}_i\bm{W}\|^2,\\
h(\bm{W})&=\sum_{l=1}^{L}\sum_{\mathcal{F}}^{\mathcal{F}\in\{s,m\}}\Gamma_{\mathcal{F},l}2Re(q_{\mathcal{F},l,l})\sqrt{\bar{\gamma}_{\mathcal{F},l}}\|\bar{\bm{h}}_l\bm{W}\|.
\end{align}
$g(\bm{W})$ and $h(\bm{W})$ are both convex functions with respect to $\bm{W}$.
To calculate the optimal $\bm{W}$ under the power constraint, we use the Difference of Convex algorithm (DCA) \cite{9964337} to handle this non-convex problem.
\textcolor{black}{Based on the local optimality conditions and duality in difference of convex programming,
the overarching method of the algorithm is to create two sequences of variables, $\bm{W}^t$, $\bm{V}^t$ so that $\bm{W}$ converges to a local optimum of the primal problem, $\bm{W}^{\star}$, and $\bm{V}$ converges to the local optimum of the dual problem, $\bm{V}^{\star}$.}
By calculating the gradient of $h(\bm{W})$ with respect to  $\bm{W}$, we can obtain $\bm{V^}t$, which is given as follows:
\begin{align}
    \frac{\partial h}{\partial\bm{W}}=\sum_{l=1}^{L}\sum_{\mathcal{F}}^{\mathcal{F}\in\{s,m\}}\Gamma_{\mathcal{F},l}Re(q_{\mathcal{F},l,l})\sqrt{\bar{\gamma}_{\mathcal{F},l}}\frac{\bm{W}^H\bar{\bm{h}}_l^H\bar{\bm{h}}_l}{\|\bar{\bm{h}}_l\bm{W}\|}\label{dh}.
\end{align}
Next, we can update the $\bm{W}^{t+1}$ by optimizing the convex problem 
\begin{align}
    \arg \min& \ \{g(\bm{W})-[h(\bm{W}^t)+<\bm{W}-\bm{W}^t,\bm{V}^t>]\}\label{DCAcon}\\
    \text{s.t.}&\ \lVert\bm{W}\rVert_F^2\leq P_{\text{max}}.\tag{\ref{DCAcon}a}\label{DCAcona}\
\end{align}
where $<\bm{X},\bm{Y}>$ denotes the inner product of $\bm{X}$ and $\bm{Y}$.
The DCA for solving Problem $\mathcal{P}\text{3-}E2$ is concluded in Algorithm 1.
\textcolor{black}{The complexity of calculating $\bm{V}^t$ in Algorithm 1 is $\mathcal{O}(LM_1^2)$.}
\textcolor{black}{And the complexity of optimizing the convex problem (\ref{DCAcon}) is $\mathcal{O}\left(\frac{M_1^4\sqrt{\log(M_1^2)}}{\epsilon^2_{DCA}}\right)$, where $\epsilon^2_{DCA}$ is the target precision of convex optimization.
The overall complexity of Algorithm 1 is $\mathcal{O}\left(t_1\left(\frac{M_1^4\sqrt{\log(M_1^2)}}{\epsilon^2_{DCA}}+LM_1^2\right)\right)$, where $t_1$ denotes the total number of iterations.}
\begin{algorithm}[t]
\caption{DCA for the Energy Matrix $\bm{W}$} 
\hspace*{0.02in} {\bf Input:} 
max number of iterations $T_{DCA,max}$, convergence error $\epsilon$, initial value of  $\bm{W}$\\
\hspace*{0.02in} {\bf Output:} optimal $\bm{W}^{\star}$
\begin{algorithmic}[1]
\STATE Initial $t=0$ and $\bm{W}^t=\bm{W}$.\\
\STATE {\bf Repeat}\\
\hspace*{0.25in}Calculate $\bm{V}^t$ according to (\ref{dh}),\\
\hspace*{0.25in}Optimize the convex problem (\ref{DCAcon}) and $t=t+1$.
\STATE {\bf Until} $\frac{\|\bm{W}^{t+1}-\bm{W}^{t}\|_F^2}{\|\bm{W}^{t}\|_F^2}\leq\epsilon$ or $T_{DCA,max}=t$.  
\STATE Output $\bm{W}^{\star}=\bm{W}^{t}$.
\end{algorithmic}
\end{algorithm}
\subsubsection{Optimization of $\bm{\theta}_1$}
With given $\bm{\Gamma},\bm{W}$, the objective has changed to optimize the phase shifts of IRS1 to improve the efficiency of energy transmission, formulated as follows:
\begin{align}
\mathcal{P}\text{4}:
\min_{\bm{W}}\sum_{l=1}^{L}\sum_{\mathcal{F}}^{\mathcal{F}\in\{s,m\}}&\Gamma_{\mathcal{F},l}e_{\mathcal{F},l}(\bm{\theta}_1)\label{P4E1}\\
\text{s.t.}\quad0\leq\theta_{1,n}<2\pi,&  \quad {n}\in\mathcal{N}_1.\tag{\ref{P4E1}{a}}\label{P4E1a}
\end{align}
which is a non-convex problem due to the constraint (\ref{P4E1a}). The similar problem has been investigated in the previous work \cite{WPCNIRS}.
Therefore, we used the algorithm in \cite{WPCNIRS} called \emph{SCA for the DL Phase Shift Optimization} to deal with this problem.
\textcolor{black}{The total complexity of the algorithm is $\mathcal{O}(t_2 L^2N_1^2)$, where $t_2$ denotes the total number of iterations that guarantees the convergence.}
By updating the variables $\bm{\phi}_1^{t+1}$ in the $(t+1)$-th iteration alternately, the convergence can be achieved. 
The details of this algorithm are omitted for brevity, which can be found in \cite{WPCNIRS}.

\section{Optimization for Transmission of Uplink Information}
Having optimized and fixed the downlink energy transmission parameters, this section focuses on the optimization of the uplink information transmission phase.
\subsection{Optimization of $\bm{\gamma}$}
Given $\bm{W},\bm{\theta}_1$ and fixed, the transmitting power of each camera is determined.
The weighted information upload latency minimization problem optimized for $\bm{\gamma}$ can be transformed into a power allocation problem in the sub-6 GHz and mmWave bands for each camera.
We can obtain the optimal power allocation factor $\gamma_l$ by optimizing the following problem
\begin{align}
\mathcal{P}\text{5}:    \max_{{\gamma}_l}&\sum_{\mathcal{F}}^{\mathcal{F}\in\{s,m\}} B_{\mathcal{F}}\log_2(1+\text{SINR}_{\mathcal{F},l})\label{P4}\\
&\text{s.t.}\quad0\leq\gamma_l\leq1,  \quad {l}\in\mathcal{L}\tag{\ref{P4}{a}}\label{P4a}.
\end{align}
Considering that the interference in the same band can vary with the $\bm{\gamma}$, we can optimize ${\gamma}_l$ device by device and the decreasing objective function value of Problem $\mathcal{P}$1 is still guaranteed.

By deriving the optimal solution for $\gamma_l$ by setting the first-order derivative of $\mathcal{P}\text{5}$ with respect to $\gamma_l$ to zero, we can obtain
\begin{align}
&B_s\frac{q_{s,l,l}}{\ln{2}\left(\sum_{i}^{L}\gamma_ip_iq_{s,l,i}+\sigma^2_s\bm{f}_{\mathcal{F},l}^H\bm{f}_{\mathcal{F},l}\right)}\notag\\
-&B_m\frac{q_{m,l,l}}{\ln{2}\left(\sum_{i}^{L}(1-\gamma_i)p_iq_{m,l,i}+\sigma^2_m\bm{f}_{\mathcal{F},l}^H\bm{f}_{\mathcal{F},l}\right)}=0\label{gamma}.
\end{align}
Similar to the optimization of $\beta$ in Section III. B, we adopt the search method based on bisection to find the optimal $\gamma^\star_l$ which satisfies (\ref{gamma}).
\subsection{Joint Optimization of $\bm{F}$ and $\bm{\theta}_2$}
Similar to the derivation of $\mathcal{P}\text{3-}E1$, the rate maximization problem can be equivalently transformed into the following MSE minimization problem
\begin{align}
\mathcal{P}\text{6}:
\min_{\bm{F},\bm{\theta}_2}\sum_{l=1}^{L}\sum_{\mathcal{F}}^{\mathcal{F}\in\{s,m\}}&(\Gamma_{\mathcal{F},l}e_{\mathcal{F},l}(\bm{F},\bm{\theta}_2)-\lambda_l\eta_lB_{\mathcal{F}}\\\notag
&-\lambda_l\eta_lB_{\mathcal{F}}\log_2({\Gamma_{\mathcal{F},l}}/{(\lambda_l\eta_lB_{\mathcal{F}}}))\label{P3_E}\\
\text{s.t.}\ (\ref{P1d}), (\ref{P1f})\tag{39a}.
\end{align}
\subsubsection{Optimization of $\bm{F}$}
With $\Gamma_{\mathcal{F},l}$ and $\bm{\theta}_2$ fixed,
$\mathcal{P}\text{6}$ is a concave function of $\bm{F}$.
we can derive the optimal solution by setting the first-order derivative of $\mathcal{P}\text{6}$ with respect to $\bm{F}$ to zero
\begin{align}
\bm{f}^{\star}_{\mathcal{F},l}=\left(\sum_{i=1}^{L}p_{\mathcal{F},i}\bar{\bm{g}}_{\mathcal{F},i}\bar{\bm{g}}_{\mathcal{F},i}^H+\sigma^2_{\mathcal{F}}\bm{I}_{M_2}\right)^{-1}\left(\sqrt{p_{\mathcal{F},l}}\bar{\bm{g}}_{\mathcal{F},l}\right).\label{optF}
\end{align}
\subsubsection{Optimization of $\bm{\theta}_2$}
With other variables all fixed and constants omitted, we can get the following equation 
\begin{align}
\sum_{l=1}^{L}\sum_{i=1}^{L}&\Gamma_{\mathcal{F},l}p_{\mathcal{F},i}\bm{f}_{\mathcal{F},l}^H\bar{\bm{g}}_{\mathcal{F},i}\bar{\bm{g}}_{\mathcal{F},i}^H\bm{f}_{\mathcal{F},l}\notag\\
=\sum_{l=1}^{L}\sum_{i=1}^{L}&\left(\Gamma_{\mathcal{F},l}p_{\mathcal{F},i}\bm{f}_{\mathcal{F},l}^H\bm{G}_{\mathcal{F}}\bm{\Phi}_2\bm{g}_{{\mathcal{F}},i}^r{\bm{g}_{{\mathcal{F}},i}^r}^H\bm{\Phi}_2^H\bm{G}_{\mathcal{F}}^H\bm{f}_{\mathcal{F},l}\right.\notag\\ 
&+\Gamma_{{\mathcal{F}},l}p_{{\mathcal{F}},i}\bm{f}_{{\mathcal{F}},l}^H\bm{g}_{{\mathcal{F}},i}^d{\bm{g}_{{\mathcal{F}},i}^r}^H\bm{\Phi}_2^H \bm{G}_{\mathcal{F}}^H\bm{f}_{{\mathcal{F}},l}\notag\\
&+\left.\Gamma_{{\mathcal{F}},l}p_{{\mathcal{F}},i}\bm{f}_{{\mathcal{F}},l}^H\bm{G}_{\mathcal{F}}\bm{\Phi}_2{\bm{g}_{{\mathcal{F}},i}^r}{\bm{g}_{{\mathcal{F}},i}^d}^H \bm{f}_{{\mathcal{F}},l}\right)\label{a},
\end{align}
\begin{align}
&\sum_{l=1}^{L}\Gamma_{{\mathcal{F}},l}\sqrt{p_{{\mathcal{F}},l}}\bar{\bm{g}}_{{\mathcal{F}},l}^H\bm{f}_{{\mathcal{F}},l}\\\notag
=&\sum_{l=1}^{L}\left(\Gamma_{{\mathcal{F}},l}\sqrt{p_{{\mathcal{F}},l}}{\bm{g}_{{\mathcal{F}},l}^d}^H\bm{f}_{{\mathcal{F}},l}+\Gamma_{{\mathcal{F}},l}\sqrt{p_{{\mathcal{F}},l}}{\bm{g}_{{\mathcal{F}},l}^r}^H\bm{\Phi}_2^H\bm{G}_{\mathcal{F}}^H\bm{f}_{{\mathcal{F}},l}\right),
\end{align}
For convenient analysis, We define all variables that are independent of $\bm{\theta}_2$ in the following form
\begin{align}
\bm{A}_{\mathcal{F}}&\triangleq\sum_{l=1}^{L}\Gamma_{{\mathcal{F}},l}\bm{G}_{\mathcal{F}}^H\bm{f}_{{\mathcal{F}},l}\bm{f}_{{\mathcal{F}},l}^H\bm{G}_{\mathcal{F}},\\ \bm{B}_{\mathcal{F}}&\triangleq\sum_{i=1}^{L}p_{{\mathcal{F}},i}{\bm{g}_{{\mathcal{F}},i}^r}{\bm{g}_{{\mathcal{F}},i}^r}^H,\label{B_s}\\
\bm{C}_{\mathcal{F}}&\triangleq\sum_{l=1}^{L}\sum_{i=1}^{L}\Gamma_{{\mathcal{F}},l}p_{{\mathcal{F}},i}{\bm{g}_{{\mathcal{F}},i}^r}{\bm{g}_{{\mathcal{F}},i}^d}^H\bm{f}_{{\mathcal{F}},l}\bm{f}_{{\mathcal{F}},l}^H\bm{G}_{\mathcal{F}}, \\ \bm{D}_{\mathcal{F}}&\triangleq\sum_{l=1}^{L}\Gamma_{{\mathcal{F}},l}\sqrt{p_{{\mathcal{F}},l}}{\bm{g}_{{\mathcal{F}},l}^r}\bm{f}_{{\mathcal{F}},l}^H\bm{G}_{\mathcal{F}}.\label{D_s}
\end{align}
Substituting (\ref{B_s})$\thicksim$(\ref{D_s}) into (\ref{e_F}) and omitting all constant, we can obtain 
\begin{align}
\sum_{l=1}^L\Gamma_{{\mathcal{F}},l}e_{{\mathcal{F}},l}=&\text{Tr}\left(\bm{\Phi}_2^H\bm{A}_{\mathcal{F}}\bm{\Phi}_2\bm{B}_{\mathcal{F}}\right)+\text{Tr}\left(\bm{\Phi}_2^H\left(\bm{C}_{\mathcal{F}}-\bm{D}_{\mathcal{F}}\right)^H\right)\notag\\
&+\text{Tr}\left(\bm{\Phi}_2\left(\bm{C}_{\mathcal{F}}-\bm{D}_{\mathcal{F}}\right)\right)\label{gbes}.
\end{align}
Therefore, by defining $\bm{E}=\sum^{s,m}_{\mathcal{F}}\bm{C}_{\mathcal{F}}-\sum^{s,m}_{\mathcal{F}}\bm{D}_{\mathcal{F}}$, we substitute (\ref{gbes}) into $\mathcal{P}\text{3-}E$. 
After omitting the constant terms, the objective function can be rewritten as
\begin{align}
\mathcal{P}\text{6-}E1:
\min_{\bm{\theta}_2}&\text{Tr}\left(\bm{\Phi}_2^H\bm{A_s}\bm{\Phi}_2\bm{B_s}\right)+\text{Tr}\left(\bm{\Phi}_2^H\bm{A_m}\bm{\Phi}_2\bm{B_m}\right)\notag\\
&+\text{Tr}\left(\bm{\Phi}_2^H\bm{E}^H\right)+\text{Tr}\left(\bm{\Phi}_2\bm{E}\right) \label{P6_E1}\\
&\text{s.t.}\ (\ref{P1d}), (\ref{P1f}) \tag{\ref{P6_E1}{a}}.
\end{align}
Upon defining $\bm{\phi}_2\triangleq\left[e^{j\theta_{2,1}},e^{j\theta_{2,2}},\dots,e^{j\theta_{2,N_2}}\right]^T$, and $\bm{e}=\left[\bm{E}_{(1,1)},\bm{E}_{(2,2)},\dots,\bm{E}_{(N_2,N_2)}\right]$, according to the properties of matrix transformation, from \cite{Matrix}, we can get
\begin{align}
&\mathrm{Tr}\left(\bm{\Phi}_2^H{\bm{A}_{\mathcal{F}}}\bm{\Phi}_2{\bm{B}_{\mathcal{F}}}\right)=\bm{\phi}_2^{H}\left({\bm{A}_{\mathcal{F}}}\odot{\bm{B}_{\mathcal{F}}}^{T}\right)\bm{\phi}_2,\label{AsodotBs}\\
&\mathrm{Tr}\left(\bm{\Phi}_2{\bm{E}}\right)={\bm{e}}\bm{\phi}_2, \ \  \mathrm{Tr}\left(\bm{\Phi}_2^{H}{\bm{E}}^{H}\right)=\bm{\phi}_2^H{\bm{e}}^{H}.
\end{align}
Accordingly, the objective function can be reformulated as
\begin{align}
\mathcal{P}\text{6-}E2:
&\min_{\bm{\theta}_2}f(\bm\theta_2)= \bm{\phi}_2^H\bm{\Xi}\bm{\phi}_2+{\bm{e}}\bm{\phi}_2+\bm{\phi}_2^H{\bm{e}}^{H}\label{P6_E2}\\
&\quad \text{s.t.}\ (\ref{P1d}), (\ref{P1f}) \tag{\ref{P6_E2}{a}},
\end{align}
where $\bm{\Xi}\triangleq{\bm{A}_s}\odot{\bm{B}_s}^{T}+{\bm{A}_m}\odot{\bm{B}_m}^{T}$.
Next, by introducing an auxiliary variable $t$ which satisfies $t^2=1$, we can get the optimization objective as follows
\begin{align}
\mathcal{P}\text{6-}E3:
\min_{\bm{\theta}_2}\quad&
\bar{\bm{\phi}}_2^H\bm{\Lambda}\bar{\bm{\phi}}_2
\label{P6_E3}\\
\text{s.t.}(\ref{P1d}), (\ref{P1f}) \tag{\ref{P6_E3}{a}},
\end{align}
where we define
\begin{align}
&\bm{\Lambda}\triangleq
\begin{bmatrix}
&\bm{\Xi} \ &\bm{e}^H\\
&\bm{e}\ &0
\end{bmatrix},
\bar{\bm{\phi}}_2\triangleq
\begin{bmatrix}
\bm{\phi}_2\\
t
\end{bmatrix}.\label{57}
\end{align}
The covert communication constraints will be analysed in the following.
From (\ref{P1f}), we can get 
\begin{align}
&\bm{d}_{\mathcal{F},k}\bm{\Phi}_2\bm{g}_{\mathcal{F},l}^r{\bm{g}_{\mathcal{F},l}^r}^H\bm{\Phi}_2^H\bm{d}_{\mathcal{F},k}^H=0, \ \mathcal{F}\in\{s,m\}, l\in\mathcal{L},\ k\in\mathcal{K}\label{convert}.
\end{align}
Then, we reformulate (\ref{convert}) as 
\begin{align}
&\bm{d}_{\mathcal{F},k}\bm{\Phi}_2\bm{g}_{\mathcal{F},l}^r{\bm{g}_{\mathcal{F},l}^r}^H\bm{\Phi}_2^H\bm{d}_{\mathcal{F},k}^H \notag\\
{\overset{(a_1)}{=}}&\text{Tr}\left(\bm{\Phi}_2^H\bm{d}_{\mathcal{F},k}^H\bm{d}_{\mathcal{F},k}\bm{\Phi}_2\bm{g}_{\mathcal{F},l}^r{\bm{g}_{\mathcal{F},l}^r}^H\right)\notag\\
{\overset{(a_2)}{=}}&\text{Tr}\left(\bm{\Phi}_2^H\bm{D}_{\mathcal{F},k}\bm{\Phi}_2\bm{G}_{\mathcal{F},l}^r\right)\notag\\
{\overset{(a_3)}{=}}&\bm{\phi}_2^{H}\left({\bm{D}_{\mathcal{F},k}}\odot{\bm{G}_{\mathcal{F},l}^r}^{T}\right)\bm{\phi}_2\notag\\
{\overset{(a_4)}{=}}&\bm{\phi}_2^{H}\bm{\Delta}_{\mathcal{F},l,k}\bm{\phi}_2,
 \ l\in\mathcal{L},\  k\in\mathcal{K}\label{convert2},
\end{align}
where ($a_1$) holds since (\ref{convert2}) is a scalar and $\text{Tr}(\bm{AB})=\text{Tr}(\bm{BA})$; 
($a_2$) holds by defining  $\bm{G}^r_{\mathcal{F},l}\triangleq\bm{g}_{\mathcal{F},l}^r{\bm{g}_{\mathcal{F},l}^r}^H$ and $\bm{D}_{\mathcal{F},k}\triangleq\bm{d}_{\mathcal{F},k}^H{\bm{d}_{\mathcal{F},k}}$ for $\mathcal{F}\in\{s,m\}$;
($a_3$) can be derived in a similar manner as (\ref{AsodotBs});
and ($a_4$) holds by defining $\bm{\Delta}_{\mathcal{F},l,k}\triangleq\left({\bm{D}_{\mathcal{F},k}}\odot{\bm{G}_{\mathcal{F},l}^r}^{T}\right)$.

Next, we apply the classical Semidefinite Relaxation (SDR) technique \cite{SDR} by defining a new auxiliary matrix $\bm{\Omega}=\bar{\bm{\phi}}_2\bar{\bm{\phi}}_2^H$.
By relaxing the highly non-convex rank-one constraint on $\bm{\Omega}$, problem P6-E3 is relaxed as follows:
\begin{align}
\mathcal{P}\text{6-}E&4:
\min_{\bm{\Omega}}\quad\text{Tr}\left(\bm{\Lambda}\bm{\Omega}\right)\label{P6_E4}\\
\text{s.t.}\quad &\bm{\Omega}_{(n,n)}=1,  \quad \text{for}  \ \forall{n}\in\left\{\mathcal{N}_2,N_2+1\right\}\tag{\ref{P6_E4}{a}}\label{P6_E4a},\\
&\bm{\Omega}\succeq0\tag{\ref{P6_E4}{b}}\label{P6_E4b},\\
&\text{Tr}\left(\Tilde{\bm{\Delta}}_{\mathcal{F},l,k}\bm{\Omega}\right)=0,\ \text{for}  \ \forall{l}\in\mathcal{L},\ \forall{k}\in\mathcal{K}.\tag{\ref{P6_E4}{c}}\label{P6_E4c}
\end{align}
where 
\begin{align}
&\Tilde{\bm{\Delta}}_{\mathcal{F},l,k}\triangleq
\begin{bmatrix}
&\bm{\Delta}_{\mathcal{F},l,k} \ &\bm{0}_{N_2\times 1}\\
&\bm{0}_{1\times N_2}\ &0
\end{bmatrix}.\label{61}
\end{align}
Consequently, problem $\mathcal{P}\text{6-}E4$ is successfully recast as a standard convex semidefinite program (SDP), which can be optimally solved using convex optimization solvers such as CVX.
To address the relaxed rank-one constraint and recover the optimal phase shift vector $\bm{\phi}_2^{\star}$ from the optimized matrix $\bm{\Omega}^{\star}$, we perform an eigenvalue decomposition on $\bm{\Omega}^{\star}$, yielding $\bm{\Omega}^{\star}=\bm{U\varSigma U}^H$.

\begin{algorithm}[t]
\caption{SDR Algorithm to Optimize $\bm{\theta}_2$} 
\hspace*{0.02in} {\bf Input:} 
$\bm{d}_{\mathcal{F},k},k=1,\dots,K,\ \bm{g}^r_{\mathcal{F},l}, l=1,\dots,L,
\bm{\Xi}, \bm{e}, N_{rand}$\\
\hspace*{0.02in} {\bf Output:} optimal $\bm{\theta}_2^{\star}$
\begin{algorithmic}[1]
\STATE Initial $\bm{\Lambda}$ and $\Tilde{\bm{\Delta}}$ according to (\ref{57}) and (\ref{61}).\\
\STATE Optimize the problem $\mathcal{P}\text{6-}E4$ to get the optimal $\bm{\Omega}^{\star}$.
\STATE Make the eigenvalue decomposition of $\bm{\Omega}^{\star}=\bm{U\varSigma U}^H$.
\STATE Generate $N_{rand}$ realizations of $\bar{\bm{\phi}}_2=\bm{U\varSigma}^{\frac{1}{2}} \bm{r}$, where $\bm{r}\sim\mathcal{CN}(0,\bm{I}_{N_2+1})$.
\STATE Select the optimal $\bar{\bm{\phi}}_2^{\star}$ by maximizing the $\bar{\bm{\phi}}_2\bm{\Lambda}\bar{\bm{\phi}}_2^H$. 
\STATE Normalize the $\bm{\phi}_2^{\star}$ as (\ref{normphi}).
\end{algorithmic}
\end{algorithm}
Next, we apply Gaussian randomization to obtain a large number of $\bar{\bm{\phi}}_2=\bm{U\varSigma}^{\frac{1}{2}} \bm{r}$, where $\bm{r}\sim\mathcal{CN}(0,\bm{I}_{M_2+1})$.
From all $\bar{\bm{\phi}}_2$ obtained with the different $\bm{r}$, we select the optimal $\bar{\bm{\phi}}_2^{\star}$
by maximizing the $\bar{\bm{\phi}}_2\bm{\Lambda}\bar{\bm{\phi}}_2^H$.
To satisfy the vector norm one constraint, we normalize the $\bm{\phi}_2^{\star}$ as follows:
\begin{align}
\bm{\phi}_2^{\star}=\exp\left({j\arg\left(\left[\frac{\bar{\bm{\phi}}_2^{\star}}{\bar{\bm{\phi}}_{2(M_2+1)}^{\star}}\right]_{1:M_2}\right)}\right).\label{normphi}
\end{align}
The overall algorithm to optimize passive phase shifts $\bm{\theta}_2$ is summarized in Algorithm 2. 
\textcolor{black}{In Step 1, the complexity of initialization   
which is dominated by calculating $\Tilde{\bm{\Delta}}$ is $\mathcal{O}(KLN_2^2+L^2N_2^2M_2)$.
In Step 2, the complexity of SDR that calculated by \emph{First-Order Methods} is $\mathcal{O}\left(\frac{N_2^2\sqrt{\log N_2}}{\epsilon_{SDR}^2}\right)$ \cite{OSDR}, where $\epsilon_{SDR}$ is the target precision of SDR.}
\textcolor{black}{Then, the complexity of Gaussian randomization from Step 3 to Step 6 is $\mathcal{O}(N_2^3+N_{rand}N_2^2)$.}
\textcolor{black}{Accordingly, the overall complexity of Algorithm 2 is $\mathcal{O}\left(N_2^2\left(KL+L^2M_2+\frac{\sqrt{\log N_2}}{\epsilon_{SDR}^2}+N_2+N_{rand}\right)\right)$.}
\subsection{Update $\bm{\lambda}$ and $\bm{\eta}$}
In the last part of the BCD algorithm, we need to optimize $\bm{\lambda}$ and $\bm{\eta}$ with all variables $\beta^{\star},\bm{W}^{\star},\bm{\theta}_1^{\star},\bm{\gamma}^{\star},\bm{F}^{\star},\bm{\theta}_2^{\star}$ optimized and fixed.
Special care needs to be taken that the problem $\mathcal{P}\text{1-}E2$ is convex programming with respect to parameters $\lambda_l\geq0$ and $\eta_l\geq0$, $l\in\mathcal{L}$. 
Accordingly, in what follows, $\bm{\lambda}$ and $\bm{\eta}$ are updated by using the modified Newton’s method \cite{MNTM} until the convergence is achieved.
Let $\bm{\mu}^t=[\bm{\lambda}^t,\bm{\eta}^t]$ denote the parameter vector in the $t$-th iteration.
According to (\ref{inipara}), we define
\begin{align}
&\psi_l^1(\bm{\mu})\triangleq-\varpi_lv_{l}+\eta_l(C^{\star}_{s,l}+C^{\star}_{m,l}),\ l\in\mathcal{L},\\
&\psi_l^2(\bm{\mu})\triangleq-1+\lambda_l(C^{\star}_{s,l}+C^{\star}_{m,l}), \ l\in\mathcal{L}.
\end{align}
Let 
\begin{align}
\psi_l(\bm{\mu})=\psi_l^1(\bm{\mu}), \quad \psi_{L+l}(\bm{\mu})=\psi_l^2(\bm{\mu}), \ l\in\mathcal{L},\label{psimu}
\end{align}
the optimal conditions $\psi_l^1(\bm{\mu})=\psi_l^2(\bm{\mu})=0, \ l=1,2,\dots,L$ can be rewritten as $\bm{\psi}(\bm{\mu})=\bm{0}$,
which means we need to update $\bm{\mu}^t$ until the convergence $\bm{\psi}(\bm{\mu}^{t+1})=0$ is recieved.
If $\bm{\psi}(\bm{\mu}^t)=0$, all variables $\beta^{\star},\bm{W}^{\star},\bm{\theta}_1^{\star},\bm{\gamma}^{\star},\bm{F}^{\star},\bm{\theta}_2^{\star}$ are the global solutions.
Otherwise, we need to iterate $\bm{\mu}^t$ by
\begin{align}
\bm{\mu}^{t+1}=\bm{\mu}^{t}-\delta^{i^{t}}[\bm{\psi}'(\bm{\mu}^{t})]^{-1}\bm{\psi}(\bm{\mu}^{t}),\label{ntm}
\end{align}
where $\delta\in(0,1)$, $\varepsilon\in(0,1)$, $i$ denotes the smallest integer among $i\in\{0,1,2,\dots\}$ satisfying
\begin{align}
\|\bm{\psi}(\bm{\mu}^{t+1})\|\leq (1-\varepsilon\delta^i)\|\bm{\psi}(\bm{\mu}^{t})\|,\label{conv}
\end{align}
and $\bm{\psi}'(\bm{\mu}^{t})$ is the Jacobian matrix of $\bm{\psi}(\bm{\mu}^{t})$, which can be easily obtained by calculating the partial derivatives as $\bm{\psi}'(\bm{\mu}^{t})=\text{diag}\left(\textbf{1}_{1\times2}\otimes\left(\bm{C}^{\star}_{s}+\bm{C}^{\star}_{m}\right)\right)$.
The $\bm{\lambda}$ and $\bm{\eta}$ are updated by iterations until $\bm{\psi}(\bm{\mu}^{t+1})=\bm{0}$ is achieved.
The whole procedures of update $\bm{\lambda}$ and $\bm{\eta}$ are summarized in Algorithm 3.
\textcolor{black}{The complexity of Algorithm 3 is dominated by calculating Jacobian matrix of $\bm{\psi}(\bm{\mu})$.}
\textcolor{black}{Since $\bm{C}_{s}$ and $\bm{C}_{m}$ are already known as the input, the complexity of Algorithm 3 can be ignored, because
all steps are given by explicit mathematical expressions.}

\begin{algorithm}[t]
\caption{Modified NewTon's Method to Update $\bm{\lambda}$ and $\bm{\eta}$} 
\hspace*{0.02in} {\bf Input:} 
$\delta, \varepsilon, \bm{\varpi}, \bm{v}, \bm{C}^{\star}_{s}, \bm{C}^{\star}_{m}, \bm{\lambda}, \bm{\eta}$\\
\hspace*{0.02in} {\bf Output:} updated $\bm{\lambda}^\text{upd}, \bm{\eta}^\text{upd}$
\begin{algorithmic}[1]
\STATE Initial $\bm{\mu}=[\bm{\lambda},\bm{\eta}]$,   $\bm{\psi}(\bm{\mu})$ according to (\ref{psimu}).\\
\STATE Update to get $\bm{\mu}^\text{upd}$ according to (\ref{ntm}), where $i$ satisfies (\ref{conv}).  
\STATE $\bm{\lambda}^\text{upd}=\bm{\mu}^\text{upd}_{(1:L)}$, $\bm{\eta}^\text{upd}=\bm{\mu}^\text{upd}_{(L+1:2L)}$.
\end{algorithmic}
\end{algorithm}
\subsection{Overall Algorithm to Solve Problem $\mathcal{P}0$}
Having completed the optimization for all individual blocks, we now summarize the overall procedure of the proposed BCD method in Algorithm 4.
Firstly, in Step 1, all variables need to be initialized, where $\beta^0\sim{U}(0,1)$, $\alpha^0_{l}\sim{U}(0,1)$, $\gamma^0_{l}\sim{U}(0,1)$, $\theta^0_{1,n}$ and $\theta^0_{2,n}\sim U(0,2\pi)$. $\bm{W}^0=\sqrt{P_{max}}\frac{\bar{\bm{W}}}{\|\bar{\bm{W}}\|_F}$, where $\bar{\bm{W}}$ is a $M_1\times M_1$ complex matrix and real and imaginary parts of each element all distribute uniformly in $[-1, 1]$. 
$\bm{F}^0$ is the same distribution as $\bar{\bm{W}}$.
$\delta\sim U(0,1)$, $\varepsilon\sim U(0,1)$.
Next, in Step 2, all variables are optimized block by block.
And in each iteration, with $F^t_{obj}\triangleq\sum_{l=1}^L\varpi_lD_l(\bm{\alpha}^t,\beta^t,\bm{W}^t,\bm{\theta}^t_1,\bm{\gamma}^t,\bm{F}^t,\bm{\theta}^t_2)$, the number of iterations and the convergence in the following should be checked to determine whether the algorithm terminates:
\begin{align}
\frac{\left|F_{obj}^t-F_{obj}^{t-1}\right|}{F_{obj}^t}\leq{\epsilon}_{2}.\label{convergence}
\end{align}

\textcolor{black}{It's worth noting that all of the proposed algorithms can guarantee to yield a monotonically decreasing objective function value of Problem $\mathcal{P}$1 in each step.
Furthermore, the objective function value has a lower bound due to the power constraint.
Accordingly, the BCD algorithm is guaranteed to converge to at least a locally optimal solution.}
\textcolor{black}{The computational complexity of Algorithm 4 is mainly
dependent on Step 2, whose complexities have been analyzed in the above subsections. 
If we fix the downlink energy transmission setting and pre-power the cameras, the complexity of the overall algorithm can be reduced to $\mathcal{O}\left(t_{out}t_{in}\left( N_2^2\left(KL+L^2M_2+\frac{\sqrt{\log N_2}}{\epsilon_{SDR}^2}+N_2+N_{rand}\right) \right)\right)$, where $t_{out}$ and $t_{in}$ are the number of outer iterations and inner iterations, respectively.}

\begin{algorithm}[t]
\caption{Overall BCD Algorithm to Solve Problem $\mathcal{P}0$} 
\textbf{Input:}
$\bm{H}, \bm{h}_l^d, \bm{h}_l^r, \bm{G}_{\mathcal{F}}, \bm{g}_{\mathcal{F},l}^r, \bm{g}_{\mathcal{F},l}^d, \bm{d}_{\mathcal{F},k}, \sigma_{\mathcal{F}}, P_{max}, T_{max}, {\epsilon}_{1}, {\epsilon}_{2}$ \\
\textbf{Output:} $\bm{\alpha}^{\star}$,$\beta^{\star}$,$\bm{W}^{\star}$,$\bm{\theta}^{\star}_1$,$\bm{\gamma}^{\star}$,$\bm{F}^{\star}$,$\bm{\theta}^{\star}_2$.
\begin{algorithmic}[1]
\STATE Initial $\bm{v}_{s}^0$,$\beta^0$,$\bm{W}^0$,$\bm{\theta}_1^0$,$\bm{\gamma}^0$,$\bm{F}^0$,$\bm{\theta}^0_2, \bm{\lambda}^0, \bm{\eta}^0, \delta, \varepsilon$ and $t=0$.\\
\STATE {\bf Repeat} (outer loop)\\
\hspace*{0.25in} $t=t+1$ and optimize $\bm{v}_{s}^t$ according to (\ref{v_sl}).\\
\hspace*{0.25in} {\bf Repeat} (inner loop)\\
\hspace*{0.25in}\hspace*{0.25in} Optimize $\beta^t$ to satisfy $\frac{\partial(f_{\mathcal{F},l})}{\partial\beta^t}=0$.\\
\hspace*{0.25in}\hspace*{0.25in} Optimize $\bm{F}_{\mathcal{F}}^t$ according to (\ref{optF}).\\
\hspace*{0.25in}\hspace*{0.25in} Optimize $\bm{W}^t$ according to \emph{Algorithm 1}.\\
\hspace*{0.25in}\hspace*{0.25in} Optimize $\bm{\theta}_1^t$ with the
\emph{SCA for the DL Phase Shift Optimization Algorithm} in \cite{WPCNIRS}.\\
\hspace*{0.25in}\hspace*{0.25in} Optimize $\bm{\gamma}^t$ to satisfy (\ref{gamma}).\\
\hspace*{0.25in}\hspace*{0.25in} Optimize $\bm{\theta}_2^t$ according to \emph{Algorithm 2}.\\
\hspace*{0.25in}\hspace*{0.25in} Update $\bm{\lambda}^t$ and $\bm{\eta}^t$ according to \emph{Algorithm 3}.\\
\hspace*{0.25in} {\bf Until} Convergence $\|\bm{\psi}(\bm{\mu}^t)\|\leq \epsilon_1$ is achieved.\\
{\bf Until} Convergence (\ref{convergence}) is achieved or $t=T_{max}$.  
\STATE $\bm{\alpha}^{\star}=[\frac{\bm{v}_{s,1}^t}{\bm{v}_1},\dots,\frac{\bm{v}_{s,L}^t}{\bm{v}_L}]$, $\beta^{\star}=\beta^t$, $\bm{W}^{\star}=\bm{W}^t$, $\bm{\theta}^{\star}_1=\bm{\theta}_1^t$, $\bm{\gamma}^{\star}=\bm{\gamma}^t$, $\bm{F}^{\star}=\bm{F}^t$, $\bm{\theta}^{\star}_2=\bm{\theta}_2^t$.
\end{algorithmic}
\end{algorithm}
\section{Trade-off between Doppler Mitigation and Latency Minimization}
In Section III and IV, we have designed and optimized the downlink and uplink settings carefully to minimize the upload latency of the proposed double IRSs aided multimedia sensing network. 
However, in HSR scenarios, the severe Doppler spread drastically reduces the channel coherence time, necessitating a prohibitive increase in pilot overhead for reliable channel estimation. 
This massive overhead severely depletes the effective system bandwidth, directly causing unacceptable transmission latency and playback stalling for bandwidth-hungry HD video streams. 
Therefore, in this section, the feasibility of mitigating the Doppler spread stemming from the movement of train is also investigated by utilizing the IRS2 based on the optimized uplink setting.
\textcolor{black}{Before presenting the specific Doppler mitigation algorithm, the well-known Doppler shift phenomenon is revisited in the proposed system.
The high-speed train travels through the cameras coverage area with a constant speed v m/s. 
For the direct link, the Doppler spread can be easily get by 
\begin{align}
f_{d,d,t}=\max(\frac{v_{speed}\cos{\varphi_{d,t}}}{\lambda_{\mathcal{F}}})=\frac{v_{speed}\cos{\varphi_{d,t}}}{\lambda_m},\label{dd}
\end{align}
where $\lambda_m$ and $\varphi_{d,t}$ are the wavelength of mmWave and horizontal angle of arrival of the direct link in $t$-th time slot, respectively.
For the cascaded link, the Doppler shift phenomenon is caused by the motion of the train leading to a change in the phase of the received signals per unit time, which results in the signals frequency change. 
Since the train and IRS2 are relatively fixed, only the cameras to the IRS2 link generates phase differences.
Without considering the near-field effect, the phase difference per unit time of the $n$-th passive element in $t$-th time slot is
\begin{align}
\Delta{\phi}_{n,t}=\frac{2\pi v_{speed}\Delta t\cos{\varphi_{c,t}}}{\lambda_m}+(\theta^{\star}_{2,n,t}-\theta^{\star}_{2,n,t-1}),\label{deltaphi}
\end{align}
where $\Delta t$ and $\varphi_{c,t}$ are one slot time and horizontal angle of arrival of the cascaded link, respectively.
Then, the Doppler spread is the maximum difference in instantaneous frequency over all significant propagation links \cite{PDC}. 
The Doppler spread of the cascaded link is
\begin{align}
f_{d,c,t}&=\max_{n_1,n_2}\left|\frac{\Delta{\phi}_{n_1,t}-\Delta{\phi}_{n_2,t}}{2\pi\Delta t}\right|\notag\\
&=\max_{n_1,n_2}\left|\frac{\left(\theta^{\star}_{2,n_1,t}-\theta^{\star}_{2,n_1,t-1}\right)-\left(\theta^{\star}_{2,n_2,t}-\theta^{\star}_{2,n_2,t-1}\right)}{2\pi\Delta t}\right|,\notag\\ &\quad\quad n_1,n_2\in\mathcal{N}_2.\label{doppler}
\end{align}
Note that $n_1$, $n_2$ refers to the different element in the IRS2.
\begin{algorithm}[t]
\caption{Heuristic Algorithm for Doppler Mitigation} 
\hspace*{0.02in} {\bf Input:} 
$\varphi_{d,t}, \varphi_{c,t}, v_{speed}, \bm{\theta^{\star}}_{2,t}, \bm{\theta^{\star}}_{2,t-1}, \lambda_{m}, N_2, po, T$, \\
\hspace*{0.02in} {\bf Output:} $\bm{\theta}^{D\!M}_{2,t}$.
\begin{algorithmic}[1]
\STATE Initial $t=1, \bm{\theta}^{D\!M}_{2,t}=\bm{\theta}^{\star}_{2,t}$.\\
\STATE {\bf Repeat}\\
\STATE\hspace*{0.25in} $t=t+1$ and calculate the Doppler spread of the direct link $f_{d,d,t}$ according to (\ref{dd}).\\
\hspace*{0.25in} {\bf for} $n=1:N_2$\\
\STATE\hspace*{0.25in}\hspace*{0.25in} Calculate $\Delta\phi_{n,t}$ according to (\ref{deltaphi}).\\
\hspace*{0.25in} {\bf end}\\
\STATE\hspace*{0.25in} Sort $\Delta\phi_{n,t}$ in descending order and update subscript index.\\
\STATE\hspace*{0.25in} Find $n\!n$, which satisfies $\frac{\Delta\phi_{n\!n,t}-{\Delta\phi_{N_2,t}}}{2\pi\Delta t}\geq f_{d,d,t}$ and $\frac{\Delta\phi_{n\!n+1,t}-{\Delta\phi_{N_2,t}}}{2\pi\Delta t}\leq f_{d,d,t}$.\\
\hspace*{0.25in} {\bf for} $n=1:n\!n$\\
\STATE\hspace*{0.25in}\hspace*{0.25in} Calculate ${\theta}^{D\!M}_{2,n,t}={\theta}^{\star}_{2,n+1,t}+{\theta}^{\star}_{2,n,t-1}-{\theta}^{\star}_{2,n+1,t-1}$.\\
\STATE\hspace*{0.25in}\hspace*{0.25in} Calculate $\mathrm{O\!B}^{D\!M}_{t}=\sum_{l=1}^{L}\varpi_lD_l({\bm{\theta}^{D\!M}_{2,t}})$ and $\mathrm{O\!B}^{\star}_{t}=\sum_{l=1}^{L}\varpi_lD_l({\bm{\theta}^{\star}_{2,t}})$.\\
\STATE\hspace*{0.25in}\hspace*{0.25in} \bf{if}  $\left(\mathrm{O\!B}_{t}^{D\!M} < \mathrm{O\!B}_{t}^{\star}\ \mathrm{or}\ \mathrm{O\!B}_{t}^{D\!M} < \mathrm{O\!B}_{t-1}^{\star}\right)$, break;\\
\hspace*{0.25in} \ \bf{end}\\
\STATE {\bf Until} $t=T$.  
\end{algorithmic}
\end{algorithm}
Mitigating Doppler spread by IRS has been proven to be feasible \cite{10606017}.
It is essential to mitigate the Doppler spread by reducing the phase difference between different paths.
Although we cannot completely eliminate the Doppler spread due to the existence of the direct link \cite{dopplermitigation2}, we can still reduce its impact on the whole system by mitigating it.
As the formula (\ref{doppler}) shows, we can optimize the phase shifts of passive elements on the IRS2 to reduce the phase difference and thus the Doppler spread.
But simply adjusting the passive elements will destroy the already aligned phase shifts.
Adjusting the phase shifts to minimize the Doppler spread will improve the coherence time of the channel but increase the information upload latency, while adjusting the phase shifts to minimize the information upload latency will result in an increase in the Doppler spread, which will also increase the information upload latency.
There is obviously a trade-off between maximization of the data rate and Doppler mitigation and how to find it is a very tricky problem.
Consequently, we proposed a low complexity heuristic algorithm which is presented in Algorithm 5, where $po$ denotes the pilot overhead and $T$ denotes the total time slots.}

\textcolor{black}{Note that in Step 6, the boundary $n\!n$ of passive elements is found since the Doppler spread of the direct link cannot be eliminated.
The Doppler spread of the cascaded link only needs to be reduced to the level of the direct link to prevent excessive performance loss.
In Step 7, we adjust the passive elements one by one to reduce phase differences.
According to (\ref{doppler}), to mitigate Doppler spread, we make ${\Delta{\phi}_{n,t}-\Delta{\phi}_{N_2,t}}={\Delta{\phi}_{n+1,t}-\Delta{\phi}_{N_2,t}}$, which means the adjusted phase shift ${\theta}^{D\!M}_{2,n,t}$ should be ${\theta}^{D\!M}_{2,n,t}={\theta}^{\star}_{2,n+1,t}+{\theta}^{\star}_{2,n,t-1}-{\theta}^{\star}_{2,n+1,t-1}$.
Then, in Step 8 and 9, the object functions are calculated to compare whether the performance degrades.
If so, the algorithm stops. }
\textcolor{black}{The complexity of the heuristic algorithm is dominated by calculating $\mathrm{O\!B}^{D\!M}$ and $\mathrm{O\!B}^{\star}$, whose complexities are on the order of $\mathcal{O}(TLM_2^2)$.}
\section{Simulations and Results}
In this section, we provide numerical results to examine the performance of our proposed algorithms.
The benefits of deploying the double IRSs in the HSR multimedia sensing network are evaluated.
We consider a HSR multimedia sensing network aided by double IRSs in single-cell for $L=3$ trackside cameras.
As for the channel model, both the small scale fading and the large scale path loss are considered.
The small scale fading follows the complex Gaussian distribution associated with zero mean and unit variance, while the path loss is given by $\rho=\rho_0\left(\frac{d}{d_0}\right)^{-\alpha}$, \textcolor{black}{where $\rho_0=-20$ dB is the large-scale fading factor at the reference distance of $d_0 = 1$ m \cite{twoscale}.
Then, we set $\alpha_{ref}=2.2$ and $\alpha_{dir}=3.5$, where $\alpha_{ref}$ denotes the path loss exponent of reflecting links and $\alpha_{dir}$ denotes that of direct links, respectively \cite{twoscale,HSRIRS5}.}
The distances from BS to IRS1 $d_H$, from BS to camera $d_{hd}$, from IRS1 to camera $d_{hr}$, from camera to IRS2 $d_{gr}$, from camera to MCR $d_{gd}$, from IRS2 to MCR $d_{G}$, from IRS2 to user $d_d$ are set as 10 m, 10 m, 5 m, 9 m, 10 m, 2 m and 7m, respectively. 
\textcolor{black}{To satisfy the zero forcing constraint in (\ref{zeroforce}),
we set the zenith of departure and zenith of arrival of the line of sight (LoS) path for all users as $\frac{3\pi}{4}$ and $\frac{\pi}{4}$, respectively.
It’s worth noting that the AoAs and AoDs of all users are still random.
This is reasonable since the geometric distribution of the passengers in the train is characterized by linearity.
With the assumption, the passive beam is easy to optimize to block users who are in the given directions.}
For all channels, the small-scale fading is assumed to be Rician fading. 
In specific, the small-scale channel can be modeled as
\begin{align}
\Tilde{\bm{H}}=\sqrt{\frac{\kappa}{\kappa+1}}\Tilde{\bm{H}}^\text{LoS}+\sqrt{\frac{1}{\kappa+1}}\Tilde{\bm{H}}^\text{NLoS},
\end{align}
where $\kappa$ is the Rician factor, $\Tilde{\bm{H}}^\text{LoS}$ is the deterministic line of sight (LoS), and $\Tilde{\bm{H}}^\text{NLoS}$ is the non-LoS (NLoS) component
\textcolor{black}{which is the Rayleigh fading.}
\textcolor{black}{The LoS component $\Tilde{\bm{H}}^\text{LoS}$ is generated by typical geometric channel model, which is given by $\Tilde{\bm{H}}^\text{LoS}=\textbf{a}_\text{xy}\left(\theta^{AoA},\theta^{ZoA}\right)\textbf{a}_\text{xy}^H\left(\theta^{AoD},\theta^{ZoD}\right)$, where $\textbf{a}_\text{xy}$ is defined as
\begin{align}
\textbf{a}_\text{xy}\left(\theta_1,\theta_2\right)=\textbf{a}_\text{x}\left(\theta_1,\theta_2\right)\otimes\textbf{a}_\text{y}\left(\theta_1,\theta_2\right),
\end{align}}\textcolor{black}{where $\textbf{a}_\text{x}$ and $\textbf{a}_\text{y}$ are antenna array response vectors, they can be written as} 
\begin{align}
\textbf{a}_\text{x}\left(\theta_1,\theta_2\right)=\Big[1,e^{(j\frac{2\pi d_\text{x}}{\lambda} \sin\theta_1 \cos\theta_2)}, \cdots, e^{(j\frac{2\pi d_\text{x}}{\lambda}(M_\text{x}-1) \sin\theta_1 \cos\theta_2)}\Big]^T,
\end{align}
\begin{align}\textbf{a}_\text{y}\left(\theta_1,\theta_2\right)=\Big[1,e^{(j\frac{2\pi d_\text{y}}{\lambda} \sin\theta_1 \sin\theta_2)},\cdots,
		e^{(j\frac{2\pi d_\text{y}}{\lambda} (M_\text{y}-1)\sin\theta_1 \sin\theta_2)}\Big]^T,\end{align} where $d_\text{x}$ and $d_\text{y}$ are the distance between elements along the X-axis and Y-axis, respectively, and $M_\text{x}$,  $M_\text{y}$ are the number of passive elements along the X-axis and Y-axis,  respectively.
{The NLoS component $\Tilde{\bm{H}}^\text{NLoS}$ follows the complex-valued Gaussian distribution that $\Tilde{\bm{H}}^\text{NLoS}\sim\mathcal{CN}(0,\bm{I})$.}

Next, the data volume of the UHD diagnostic video chunk $v_l$ generated within the buffering window is assumed to follow a uniform distribution, i.e., $v_l \sim U(1000, 2000)$ kb.
The weights of all cameras are also obey uniform distribution.
To facilitate analysis, the volumes of all tasks and weights are fixed in the following simulations.
\textcolor{black}{The other default settings of system-related parameters are summarised as follows: Antenna numbers $M_1\ /\ M_2\ $ = 25\ / \ 9. Passive element numbers $N_1\ /\ N_2\ $ = 100\ / \ 100. Number of users and cameras $K$ and $L$ are all 3. Carrier frequency $f_s/f_m$ = 3.5 \ /\ 28 GHz \cite{hfb1}. Bandwidths $B_s/B_m$ = $10$\ /\ 80 MHz \cite{hybridfre}. Rician factor $\kappa$ = 6 \cite{litianyou2}. Transmission power  $P_{\text{max}}$ = 10 W. Power spectral density of the noise = -174 dBm \cite{hybridfre}.}

\begin{figure}{}
\centering
	\includegraphics[width=0.4\textwidth,height=0.23\textheight]{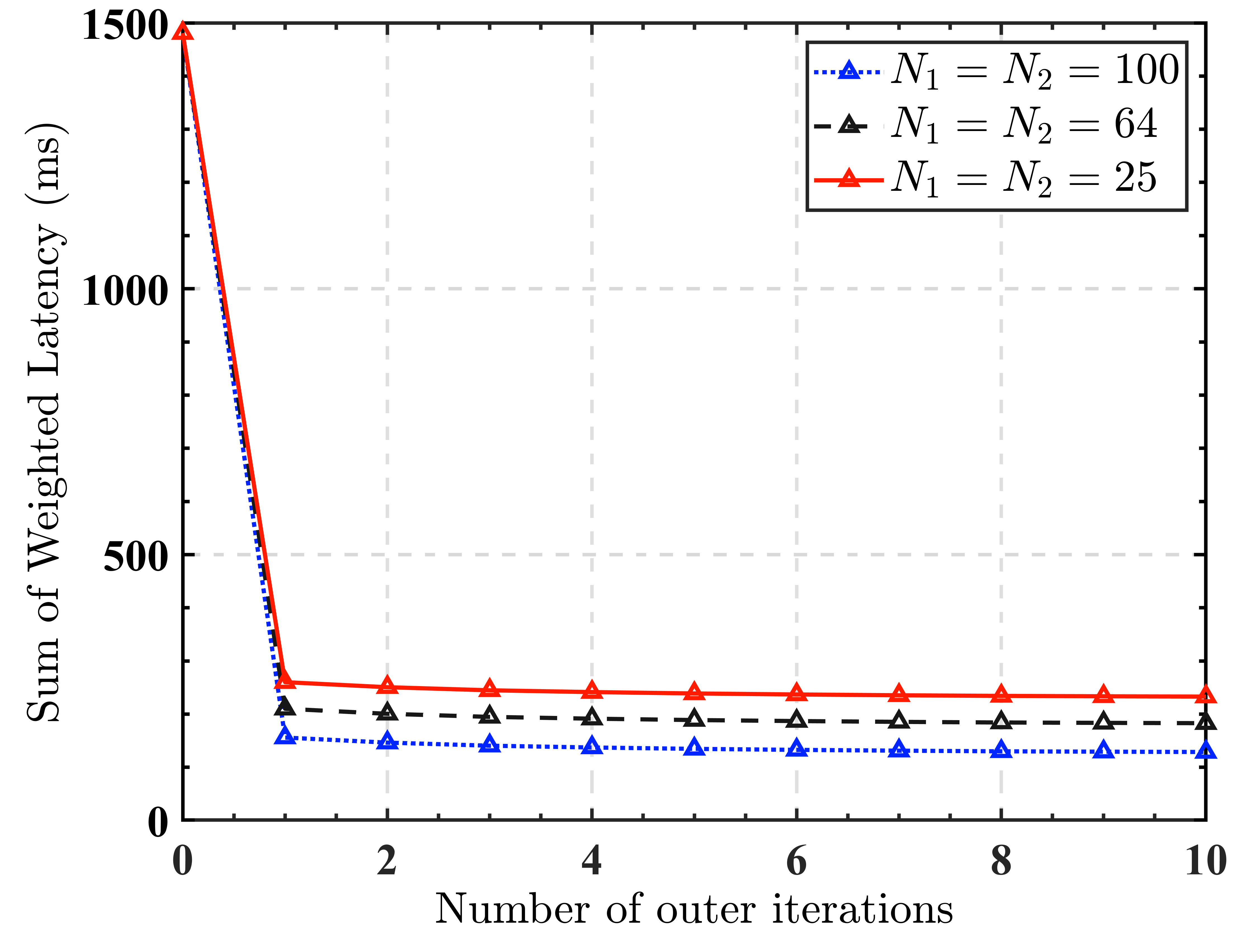}
	\caption{Outer iterations convergence behavior of the BCD Algorithm.
}\label{fig3}
\end{figure}
\begin{figure*}
\centering
	\includegraphics[width=0.87\textwidth,height=0.18\textheight]{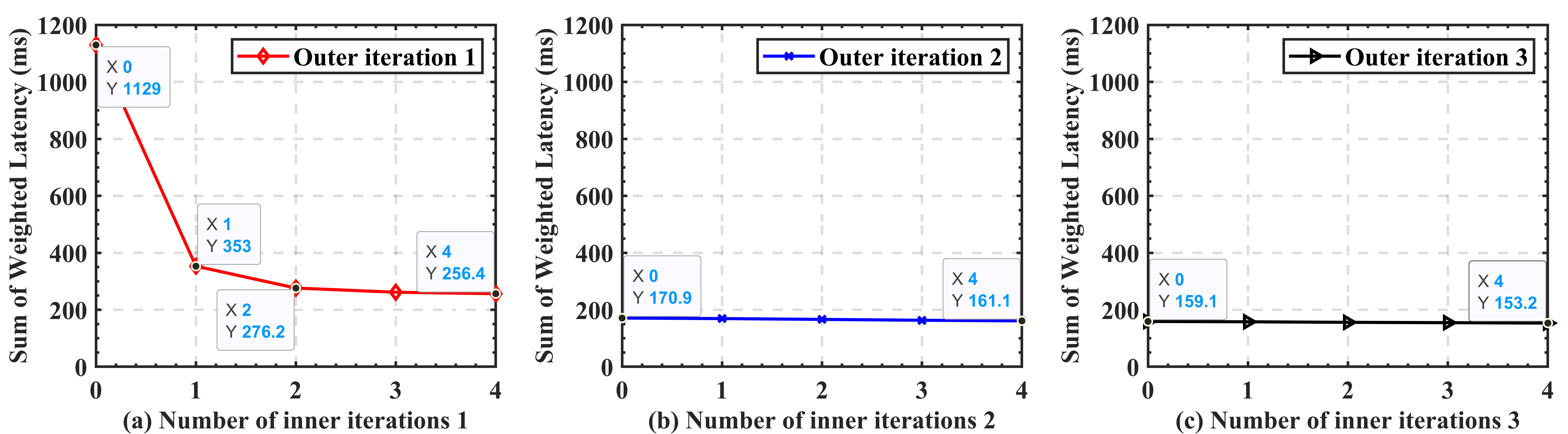}
	\caption{Sum of weighted latency versus the number of inner iterations}\label{fig4}
\end{figure*}

\begin{table}[]
\centering
\caption{Simulation Parameters}
\begin{tabular}{|c|c|lll}
\cline{1-2}
Antenna numbers $M_1\ /\ M_2\ $ & 25\ / \ 9 & \\ \cline{1-2}
Passive element numbers $N_1\ /\ N_2\ $ & 100\ / \ 100 & \\ \cline{1-2}
Number of passengers $K$& 3  & \\ \cline{1-2}
Number of cameras $L$& 3  & \\ \cline{1-2}
Carrier frequency $f_s/f_m$ & 3.5 GHz\ /\ 28 GHz \cite{hfb1}& \\ \cline{1-2}
Bandwidths $B_s/B_m$    &  $10$\ /\ 80 MHz  \cite{hybridfre}      &      \\ \cline{1-2}
Rician factor $\kappa$&  6    \cite{litianyou2}    &      \\ \cline{1-2}

Transmission power  $P_{\text{max}}$    &    10 W      &      \\ \cline{1-2}
Power spectral density of the noise       &  -174 dBm    \cite{hybridfre}    &      \\ \cline{1-2}
Energy transmission efficiency $\xi$&  0.8        &      \\ \cline{1-2}
\end{tabular}
\end{table}
First, we simulate the convergence behavior of our proposed BCD Algorithm averaged over 100 independent channel realizations
and the result is shown in Fig. 2 and Fig. 3.
It’s observed that our proposed BCD algorithm converges monotonically, which is in line with our analysis in Section IV. 
In Fig. 2, the outer loop can decrease quickly in two iterations.
The sum of weighted latency drops by $90\%$ in the first iteration, which shows that we can obtain most of the latency improvement in the first two iterations and our proposed algorithm has an extremely low complexity.
What's more, the convergence behaviors show similar trends under different settings of double IRSs phase shift number, i.e. $N_1=N_2=25,\ 64,$ and $100$.
As the number of elements increases, the latency decreases.
The reason is obvious that the energy efficiency of the transmission and the channel capacity are enhanced.
Although the BCD algorithm may require more than 10 outer iterations for full convergence, the latter iterations have little performance improvement, and in a practical scenario, we can set the maximum number of outer iterations to 2.
In Fig. 3, the inner loop behaviors are investigated.
From Fig. 3(a), the latency can almost converge in three inner iterations of the first outer iteration.
And in Fig. 3(b), the latency can obtain little improvement in three inner iterations of the second outer iteration.
A similar conclusion can be drawn from Fig. 3(c).


Next, we compare the BCD Algorithm with the following benchmark schemes:

\textcolor{black}{1) \textbf{Random Energy Beams $\bm{W}$}: The energy beams $\bm{W}$ in downlink energy transmit setting is fixed in each iteration. Both the real and imaginary parts of the energy beams matrix $\bm{W}$ follow independent uniform distribution in (0, 1). Then, it is initialized by $\lVert\bm{W}\rVert_F^2= 10$. The numbers of passive elements of both IRSs are all fixed at 100.}

2) \textbf{No IRS1/IRS2}: The system without one of double IRSs still can complete the transmission task, though there will be some performance loss. The number of passive elements of the other one IRS is fixed at 100.

3) \textbf{Random Phase Shifts with IRS1/IRS2}: The phase shifts of one IRS are generated randomly following uniform distribution in [0, 2$\pi$), while the phase shifts of the other one are optimized. Similarly, the number of passive elements of the other IRS is fixed at 100.
\begin{figure}{}
\centering
	\includegraphics[width=0.4\textwidth,height=0.23\textheight]{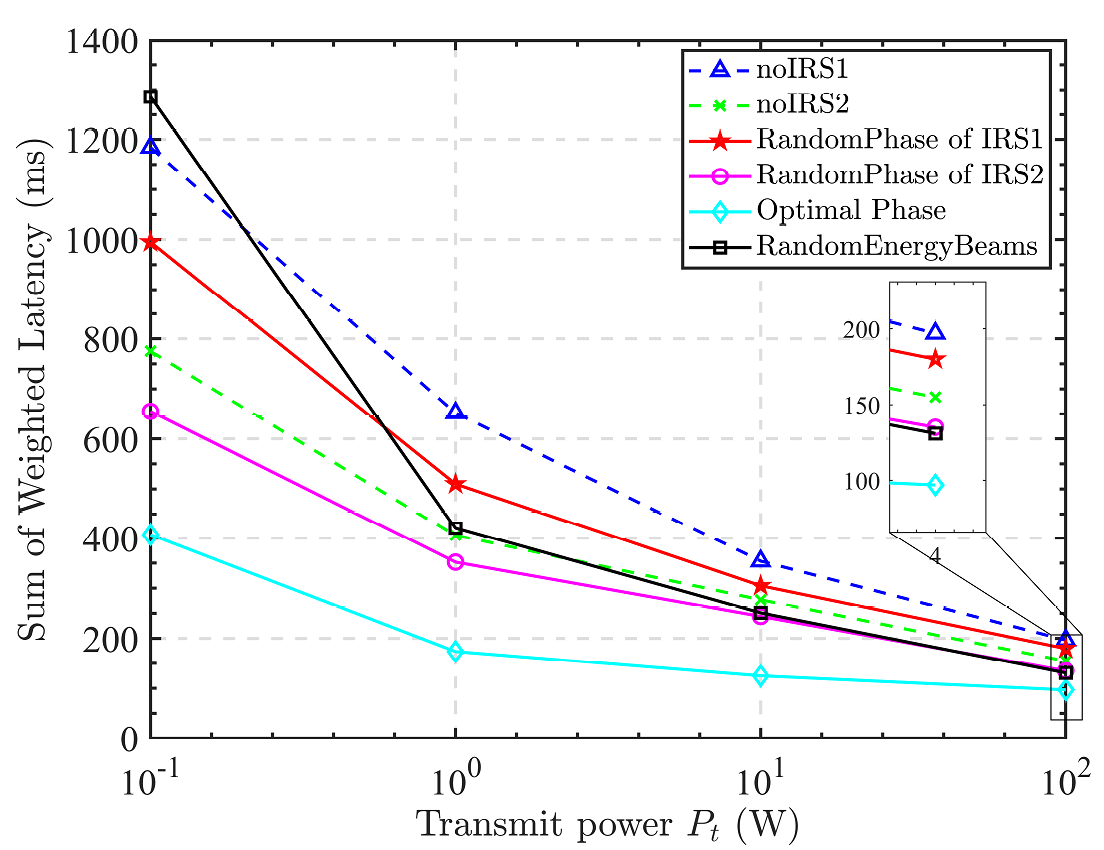}
	\caption{\textcolor{black}{Sum of weighted latency versus the transmit power $P_t$.}}\label{fig5}
\end{figure}
In Fig. 4, we compare the latency by all schemes versus
the transmit power $P_t$ of the BS. 
Specifically, the latency reductions of all schemes become quite pronounced upon increasing transmit power.
This is obvious because the cameras can harvest more energy from the BS for information \textcolor{black}{transmission}.
\textcolor{black}{And when the random energy beams $\bm{W}$ is fixed without optimization, there is a significant drop in latency performance, especially at low transmitting power.
As the transmit power rises, the sum weighted latency gap between the RandomEnergy Beams  and the Optimal Phase decreases, so that the optimization of $\bm{W}$ has a greater impact on the system when energy is scarce.}

The optimized phase shifts of double IRSs can efficiently decrease the latency of information upload than that of random phase shifts in benchmarks.
Additionally, our proposed algorithm performs better in low transmit power condition, which means the double IRSs deployment scheme is more suitable for energy-scarce scenarios.
This is mainly because when transmit power is too low, the performance bottleneck of the whole system is the energy received by all cameras. 

Fig. 5 presents the influence of performance latency for the five benchmark schemes and the proposed BCD algorithm versus the number of passive elements.
Firstly, similar to what we learned from Fig. 4, our proposed BCD algorithm obviously outperforms other benchmarks.
The latency of all schemes \textcolor{black}{decreases} with the number of passive elements.
The sum of weighted latency can decrease to 130 ms.
Secondly, as we can see \textcolor{black}{from} curves of both IRSs, the performance gap between the benchmarks “noIRS1/IRS2” and “RandomPhase of IRS1/IRS2” becomes higher upon increasing the number of reflecting elements.
It implies that the double IRSs are capable of assisting the energy and information transmission.
And with careful phase shifts optimized by the proposed algorithm, the performance can be further improved.
Thirdly, under the default system setting, \textcolor{black}{deploying} the IRS at the camera side to assist the energy transmission performs better than that at the MCR side to assist the information transmission.
We will discuss how to allocate the elements of IRSs in the next.

\textcolor{black}{In Fig. 6, we study the impact of allocating passive elements of double IRSs on the system latency. 
The curve changes versus the number of passive elements on IRS1,} with the total number of elements is fixed at 200, which means $N_1 + N_2 = 200$.
It's obvious that with the transmit power increases, the summation of weighted latency decreases.
Furthermore, as a resource, a certain number of passive elements should be appropriately allocated for maximization of the harvest energy and channel capacity to minimize the latency. 
Accordingly, there exists an inherent trade-off in the $N_1 / N_2$ allocation among the double IRSs. 
Note that the elements allocation is more significant in low transmitting power case.
\textcolor{black}{As the transmit power increases, the effect of elements allocation becomes weaker and weaker.
Although the increase in power can lead to a decrease in latency, it is not linear. 
This is a typical phenomenon of diminishing marginal benefits. The optimal elements assignment reduces the latency by approximately $50\%$ compared to assigning the elements all to one IRS.
So $N_1$ and $N_2$ require more careful design especially when energy is scarce.
Furthermore,} when energy transmission power $P_t=1W$, the latency of the scheme can be minimized at approximately $N_1 = 100$, $N_2 = 100$.
With the increase of $P_t$, the trade-off point moves to the right.
The main reason is that when energy transmission is low, there is a performance bottleneck due to received energy by all cameras.
\textcolor{black}{Even though the IRS2 improves the channel conditions, the lack of transmission energy leads to performance degradation.
In the case of lower transmitting power, more elements that deployed on IRS1 can help achieve the desirable latency.}
And the specific elements of IRSs allocation scheme needs to be determined according to the specific environment.

In the next, we define the total residual co-channel interference power to the MBMS passengers as follows:
\begin{align}
\zeta=\sum_{k=1}^{K}\sum_{l=1}^{L}\sum_{\mathcal{F}}\left|\bm{d}_{\mathcal{F},k}\bm{\Phi}_2\right.&\left.\bm{g}_{{\mathcal{F}},l}^r\right|^2,
\end{align}
  \begin{figure}{}
\centering
	\includegraphics[width=0.4\textwidth,height=0.23\textheight]{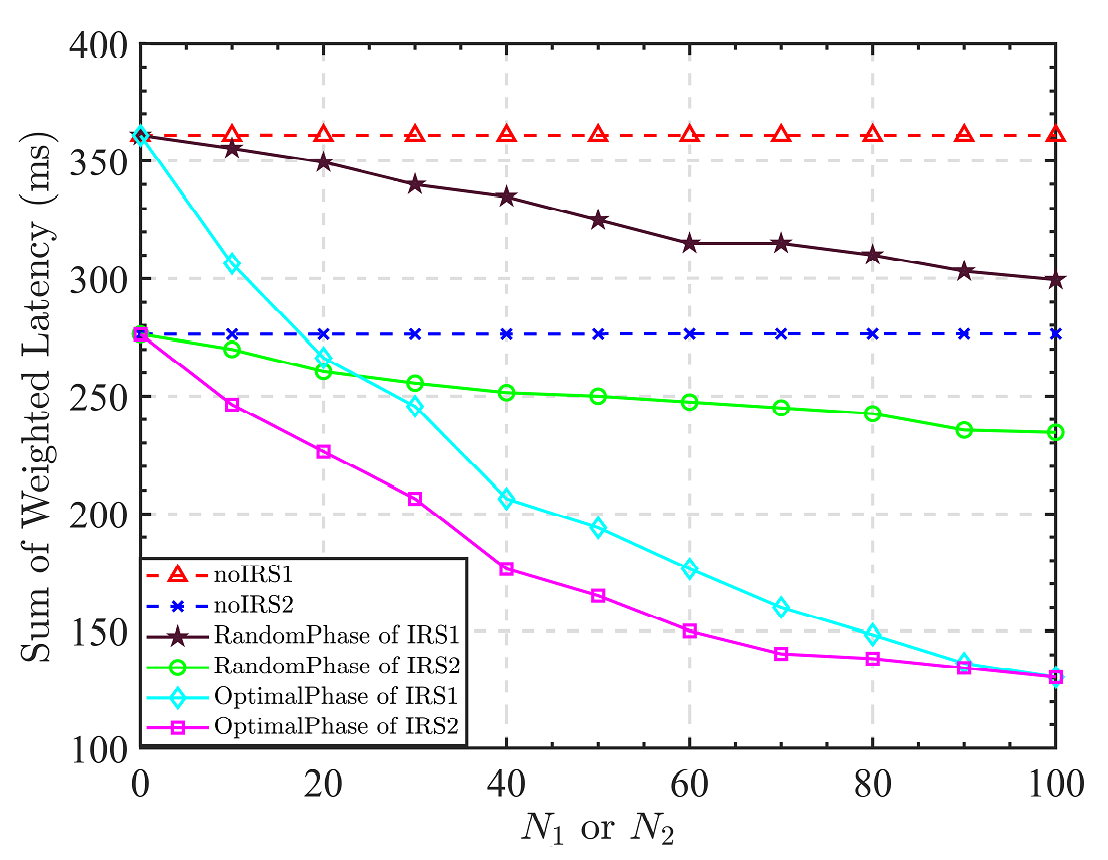}
	\caption{\textcolor{black}{Sum of weighted latency versus the number of passive elements.}}\label{fig6}
\end{figure}
\begin{figure}{}
\centering
	\includegraphics[width=0.4\textwidth,height=0.23\textheight]{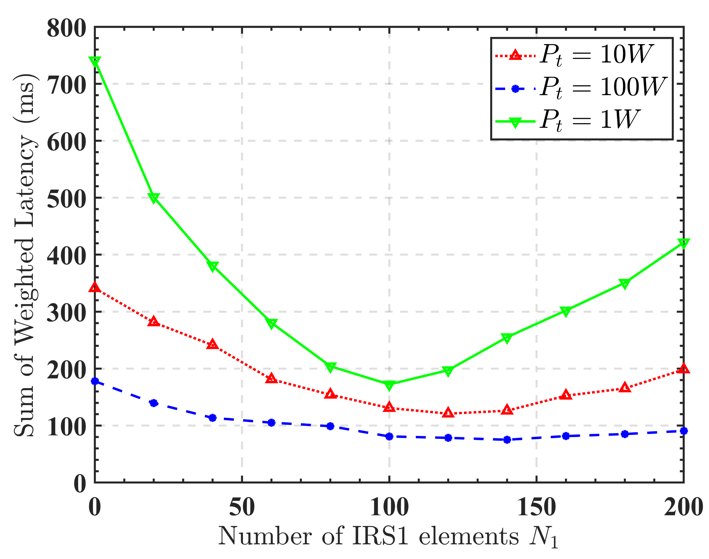}
	\caption{Sum of weighted latency versus the number of passive elements $N_1$ with $N_1 + N_2 = 200$. (b) Sum of weighted latency versus the number of passive elements.}\label{fig7}
\end{figure}
\begin{figure}{}
\centering
	\includegraphics[width=0.4\textwidth,height=0.23\textheight]{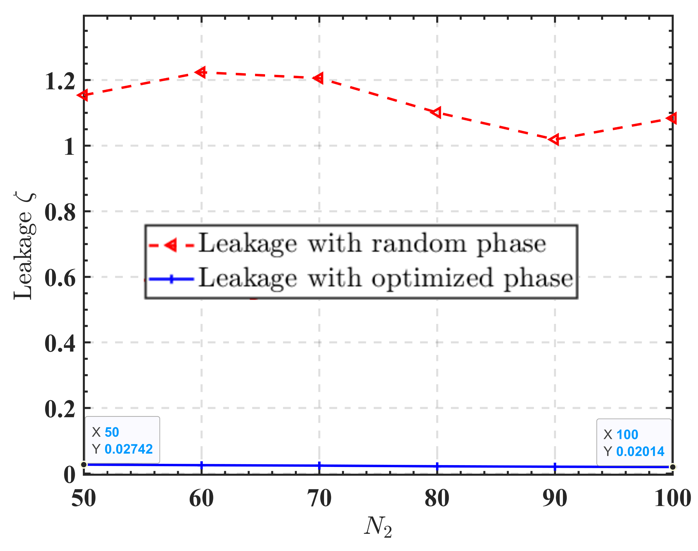}
	\caption{Sum of weighted latency versus the number of passive elements.}\label{fig8}
\end{figure}
As shown in Fig. 7, we investigate $\zeta$ against the number of passive elements of IRS2. 
From the figure, with the increase of the number of passive elements, the interference power $\zeta$ with optimized phase shifts does not change significantly and never drops to exactly 0, though the constraint (\ref{P6_E4c}) is introduced.
The main reason is that the passive beamforming with finely designed of IRS2 can shield the LoS path to the users, but some of other NLoS paths will still cause slight interference leakage.
Also, we relaxed the rank one constraint when optimizing the phase shifts in Algorithm 2, which means the strict zero-forcing constraint (\ref{P6_E4c}) has been slightly compromised when eigenvalue decomposition and Gaussian randomization are applied.
Nevertheless, the total residual interference is reduced by approximately 97.5\% compared to the pre-optimization period.
With the deployment of the IRS2, the total leakages to all users can be reduced to an extremely low level.
Although the improvement is slight, the system exhibits a decrease in leakage as the number of passive elements increases.
This is because with more elements, the narrower passive beams are generated and the leakage to NLoS paths decreases.
\textcolor{black}{It is worth noting that a sufficient number of passive elements need to be deployed to ensure that the passive beam can be finely designed, otherwise problem $\mathcal{P}\text{6-}E4$ will lead to no solution due to constraint (\ref{P6_E4c}).}
In conclusion, the proposed multimedia sensing network aided by IRSs achieves excellent spatial interference isolation within the carriage from the perspective of the physical layer.

Then, to validate the effectiveness of the proposed Doppler mitigation (DM) algorithm, we evaluate the sum of weighted latency against the train's position as it passes through the trackside diagnostic zone (e.g., TFDS inspection checkpoint) where the UHD cameras and IRS1 are deployed.
We consider a dynamic simulation setting where the high-speed train approaches, passes, and subsequently departs from the trackside cameras.
The train is 20 m away from the camera at the farthest and only 1 m at the closest.
And the pilot overhead ($po$) is fixed at 6000.
From the Fig. 8, as the train travels, the sum weighted latency of all curves is constantly changing. 
What's more, the sum weighted latency of the system is directly related to the distance between the cameras and the train.  
It is higher when the train is further away from the cameras.
Due to the influence of channel coherence time, the faster the train's speed, the worse the system performance.
More importantly, our proposed DM algorithm can greatly improve the performance of communication system than no DM algorithm and thus reduce the sum weighted latency.
The performance improvement of the proposed algorithm is more significant when the train's speed is faster.
At the farthest distances, the sum weighted latency can be reduced by 40$\%$ when train's speed v=110 m/s while there is little improvement when v = 50 m/s.
It's worth noting that as the train's speed increases, the applicable distance of the DM algorithm becomes longer and longer, from 0$\sim$4 m at v = 50 m/s to 0$\sim$15 m at v = 110m/s.
This is obvious because as the distance gets farther and the speed gets faster, the greater the Doppler spread, and the greater the Doppler mitigation by the proposed algorithm.
Then, the train's speed is fixed at v = 70 m/s ($\approx$250 km/h).
The simulation result is shown in Fig. 9.
\begin{figure}{}
\centering
	\includegraphics[width=0.4\textwidth,height=0.23\textheight]{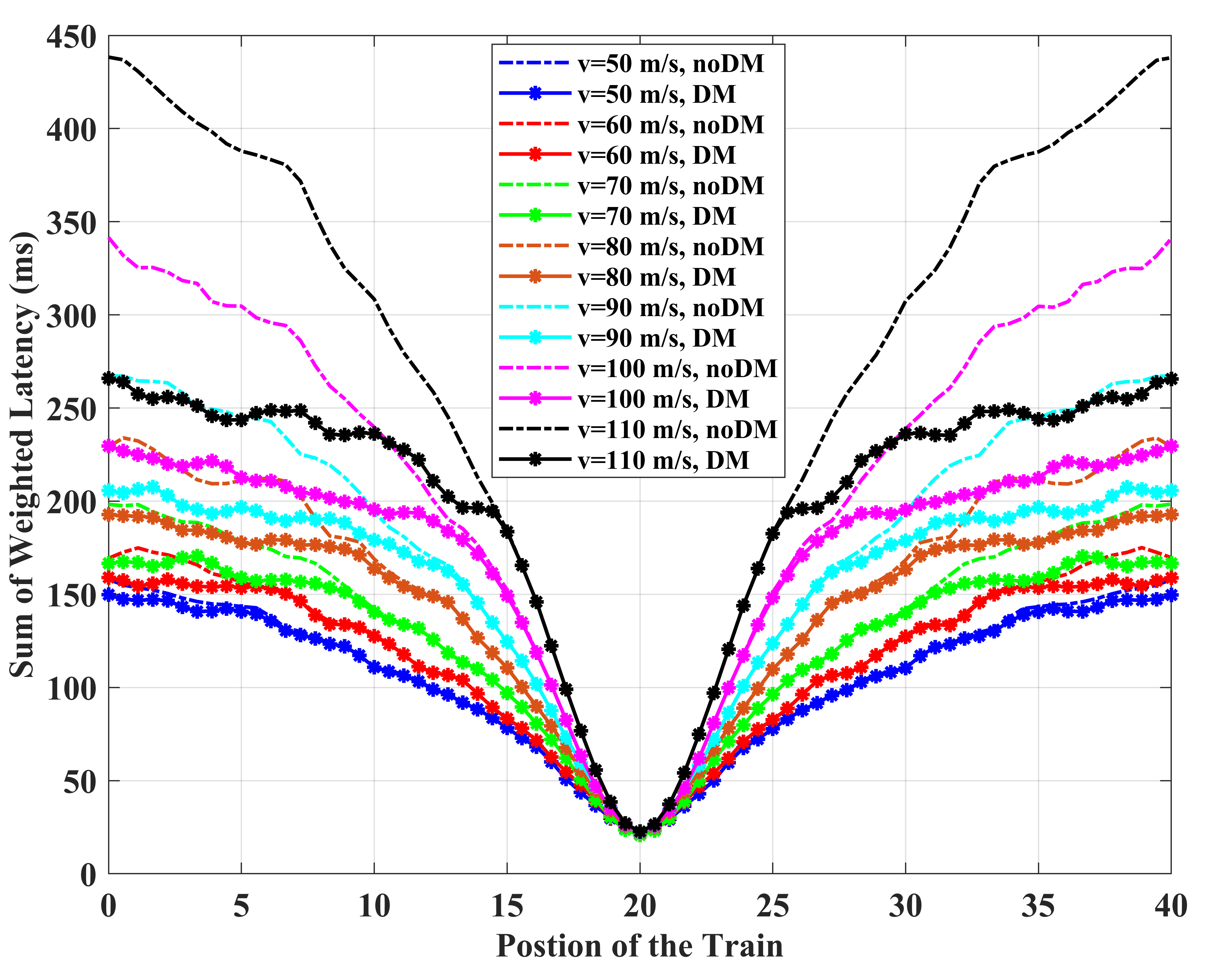}
	\textcolor{black}{\caption{Sum of weighted latency versus train's position with different speed.}\label{fig9}}
\end{figure}
\begin{figure}{}
\centering
	\includegraphics[width=0.4\textwidth,height=0.23\textheight]{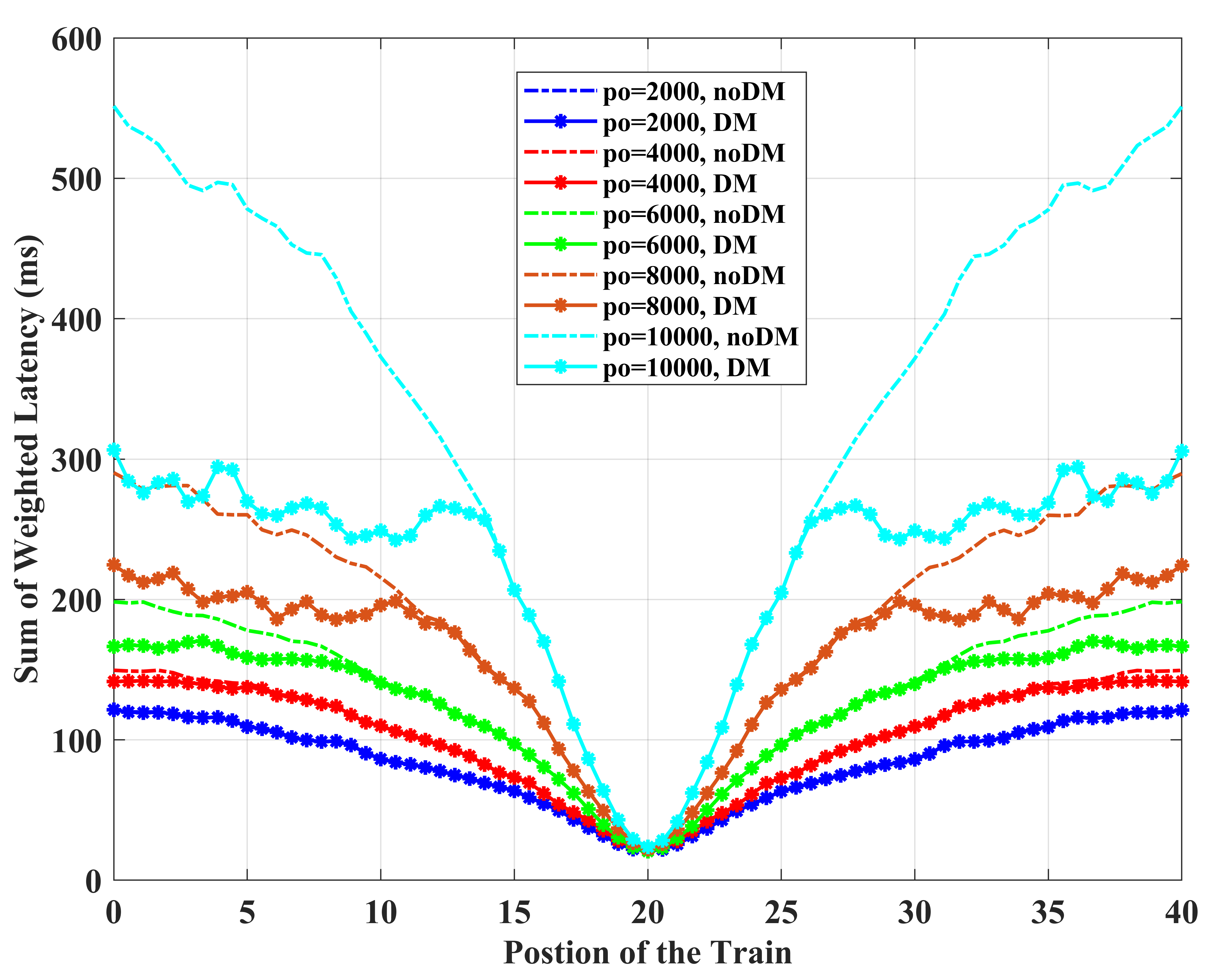}
	\textcolor{black}{\caption{Sum of weighted latency versus the position of train with different pilot overhead.}\label{fig10}}
\end{figure}
Similar to that in Fig. 8, when $po$ is higher, the performance improvement is more significant.
\section{Conclusion}
In this paper, we proposed a novel 6G I2V communication framework for HSR mechanical fault diagnosis by deploying double IRSs at both the trackside and the train windows. 
Our simulation results verified that by carefully designing the phase shifts of the double IRSs block by block, the sum of information upload latency can be minimized by more than 90\%, strictly satisfying the URLLC thresholds for 4K video bursts. 
Furthermore, the window-mounted IRS successfully functions as a spatial filter, enabling an extreme suppression of co-channel interference to the in-cabin MBMS users. 
Finally, the inherent trade-off in the passive element allocation between the two IRSs was revealed, indicating that element deployment must be strategically customized according to specific energy constraints.
In our future work, we will consider employing a deep neural network that takes easily obtainable train state characteristics—such as real-time position and velocity—as inputs to generate the optimal historical hybrid beams.

\ifCLASSOPTIONcaptionsoff
  \newpage
\fi



%
\footnotesize
\bibliographystyle{IEEEtran}
\bibliography{Bibliography}

\end{document}